\documentclass[journal,twoside,web]{ieeecolor}
\usepackage{jsen}
\usepackage{amsmath,amssymb,amsfonts}
\usepackage{algorithmic}
\usepackage{graphicx}
\usepackage{textcomp}
\usepackage{wrapfig}
\usepackage{tabularx}
\usepackage{makecell}
\usepackage{eurosym}

\usepackage{subcaption}

\usepackage[shortcuts,acronym]{glossaries} 
\usepackage[shortcuts,acronym]{glossaries}

\newacronym{uwb}{UWB}{ultra-wideband}
\newacronym{ir-uwb}{IR-UWB}{impulse-radio ultra-wideband}
\newacronym{fmcw}{FMCW}{frequency-modulated continuous wave}
\newacronym{ble}{BLE}{bluetooth low energy}
\newacronym{mmwave}{mmWave}{millimeterwave}
\newacronym{gnss}{GNSS}{global navigation satellite system}

\newacronym{fr3}{FR3}{frequency range 3}
\newacronym{ch5}{CH5}{channel 5}
\newacronym{ch9}{CH9}{channel 9}

\newacronym{jcas}{JCAS}{joint communication and sensing}
\newacronym{isac}{ISAC}{integrated sensing and communication}

\newacronym{ioe}{IoE}{Internet of Everything}
\newacronym{iot}{IoT}{Internet of Things}
\newacronym{iiot}{IIoT}{Industrial Internet of Things}

\newacronym{psd}{PSD}{Power Spectral Density}

\newacronym{afsiw}{AFSIW}{Air-Filled Substrate-Integrated-Waveguide}
\newacronym{cbs}{CBS}{Cavity-Backed Slot}
\newacronym{dra}{DRA}{Dielectric Resonator Antenna}
\newacronym{mimo}{MIMO}{Multiple Input Multiple Output}
\newacronym{hpbw}{HPBW}{Half-Power Beamwidth}
\newacronym{ftbr}{FTBR}{Front-to-Back Ratio}
\newacronym{sff}{SFF}{System Fidelity Factor}
\newacronym{dee}{DEE}{Distance Estimation Error}
\newacronym{fov}{FOV}{field-of-view}
\newacronym{qm}{QM}{Quarter-Mode}

\newacronym{rtls}{RTLS}{real-time location system}
\newacronym{tof}{ToF}{Time of Flight}
\newacronym{rtt}{RTT}{Round Trip Time}
\newacronym{toa}{ToA}{Time of Arrival}
\newacronym{twr}{TWR}{Two-way Ranging}
\newacronym{tdoa}{TDoA}{Time Difference of Arrival}
\newacronym{pdoa}{PDoA}{Phase Difference of Arrival}
\newacronym{aoa}{AoA}{Angle of Arrival}
\newacronym{acir}{ACIR}{accumulated channel impulse response}
\newacronym{cir}{CIR}{channel impulse response}
\newacronym{dop}{DOP}{dilution of precision}

\newacronym{mae}{MAE}{mean absolute error}
\newacronym{rmse}{RMSE}{root mean square Error}

\newacronym{music}{MUSIC}{Multiple Signal Classification}
\newacronym{fft}{FFT}{Fast Fourier Transform}
\newacronym{ekf}{EKF}{Extended Kalman Filter}
\newacronym{iq}{IQ}{in-phase and quadrature}

\newacronym{gpu}{GPU}{graphics processing unit}
\newacronym{fpu}{FPU}{floating-point unit}
\newacronym{dat}{DAT}{dual-antenna tile}
\newacronym{cots}{COTS}{commercial off-the-shelf}
\newacronym{trx}{TRX}{transceiver}
\newacronym{ic}{IC}{integrated circuit}
\newacronym{pcb}{PCB}{Printed Circuit Board}
\newacronym{tcxo}{TCXO}{temperature-controlled crystal oscillator}
\newacronym{pu}{PU}{processing unit}
\newacronym{pmic}{PMIC}{power management integrated circuit}

\newacronym{spi}{SPI}{Serial Peripheral Interface}
\newacronym{i2c}{I2C}{Inter-Integrated Circuit}
\newacronym{csn}{CSn}{Chip Select}
\newacronym{sda}{SDA}{Serial Data}
\newacronym{scl}{SCL}{Serial Clock}
\newacronym{gpio}{GPIO}{general-purpose input-output}
\newacronym{io}{IO}{input-output}

\newacronym{rtos}{RTOS}{real-time operating system}
\newacronym{soc}{SoC}{system-on-chip}
\newacronym{rf}{RF}{radio frequency}
\newacronym{gcpw}{GCPW}{grounded coplanar waveguide}
\newacronym{tx}{TX}{transmit}
\newacronym{rx}{RX}{receive}

\newacronym{los}{LOS}{line-of-sight}
\newacronym{nlos}{NLOS}{non-line-of-sight}

\newacronym{roi}{ROI}{region of interest}

\newacronym{dk}{DK}{development kit}
\newacronym{sram}{SRAM}{static random access memory}

\newacronym{fp}{FP}{first path}

\newacronym{ml}{ML}{machine learning}
\newacronym{ai}{AI}{artificial intelligence}
\newacronym{cnn}{CNN}{convolutional neural network}
\newacronym{lstm}{LSTM}{long short-term memory}
\newacronym{dnn}{DNN}{deep neural network}
\newacronym{nn}{NN}{neural network}

\newacronym{rr}{RR}{respiration rate}
\newacronym{br}{BR}{breathing rate}
\newacronym{hr}{HR}{heart rate}
\newacronym{rbm}{RBM}{random body movement}
\newacronym{bpm}{BPM}{breaths per minute}

\newacronym{pll}{PLL}{phase locked loop}
\newacronym{shr}{SHR}{synchronization header}
\newacronym{phr}{PHR}{physical layer header}
\newacronym{psdu}{PSDU}{physical layer service data unit}

\newacronym{mac}{MAC}{medium access control}
\newacronym{rs}{RS}{reed solomon}

\usepackage{cleveref} 
\crefname{table}{table}{tables}
\Crefname{table}{Table}{Tables}
\crefname{figure}{fig.}{Figures}
\Crefname{figure}{Fig.}{Figures}
\Crefname{section}{Section}{Sections}

\usepackage{pifont}

\usepackage{url}

\usepackage[authormarkup=none, final]{changes}
\definechangesauthor[name={rev1}, color=blue]{rev1}
\definechangesauthor[name={rev2}, color=orange]{rev2}
\definecolor{darkgreen}{HTML}{2E7D32}
\definechangesauthor[name={rev3}, color=darkgreen]{rev3}

\def\BibTeX{{\rm B\kern-.05em{\sc i\kern-.025em b}\kern-.08em
    T\kern-.1667em\lower.7ex\hbox{E}\kern-.125emX}}
\definecolor{abstractbg}{rgb}{0.89804,0.94510,0.83137}
\definecolor{dark_green}{rgb}{0, 0.6705, 0.180}
\newcommand{\cmark}{\ding{51}}%
\newcommand{\xmark}{-}%

\begin{document}
\title{Low-cost Embedded Breathing Rate Determination Using 802.15.4z IR-UWB Hardware for Remote Healthcare}
\author{Anton Lambrecht, Stijn Luchie, Jaron Fontaine, Ben Van Herbruggen, Adnan Shahid and Eli De Poorter
\thanks{Received 21 February 2026; The research that led to these results has received funding by the DistriMuSe project (HORIZON-KDT-JU-2023-2-RIA) with Grant No 101139769, from the Belgian Defense through contract number 24DEFRA009 and from the FWO PESSO project with Grant No 3G018522.}
\thanks{The authors are affiliated with IDLab, Department of Information Technology, Ghent University-imec, iGent Tower, Technologiepark-Zwijnaarde 126, B-9052 Ghent, Belgium. (email: Anton.Lambrecht@Ugent.be)}}

\IEEEtitleabstractindextext{%
\fcolorbox{abstractbg}{abstractbg}{%
\begin{minipage}{\textwidth}%
\begin{wrapfigure}[16]{c}{4in}%
\includegraphics[width=3.9in]{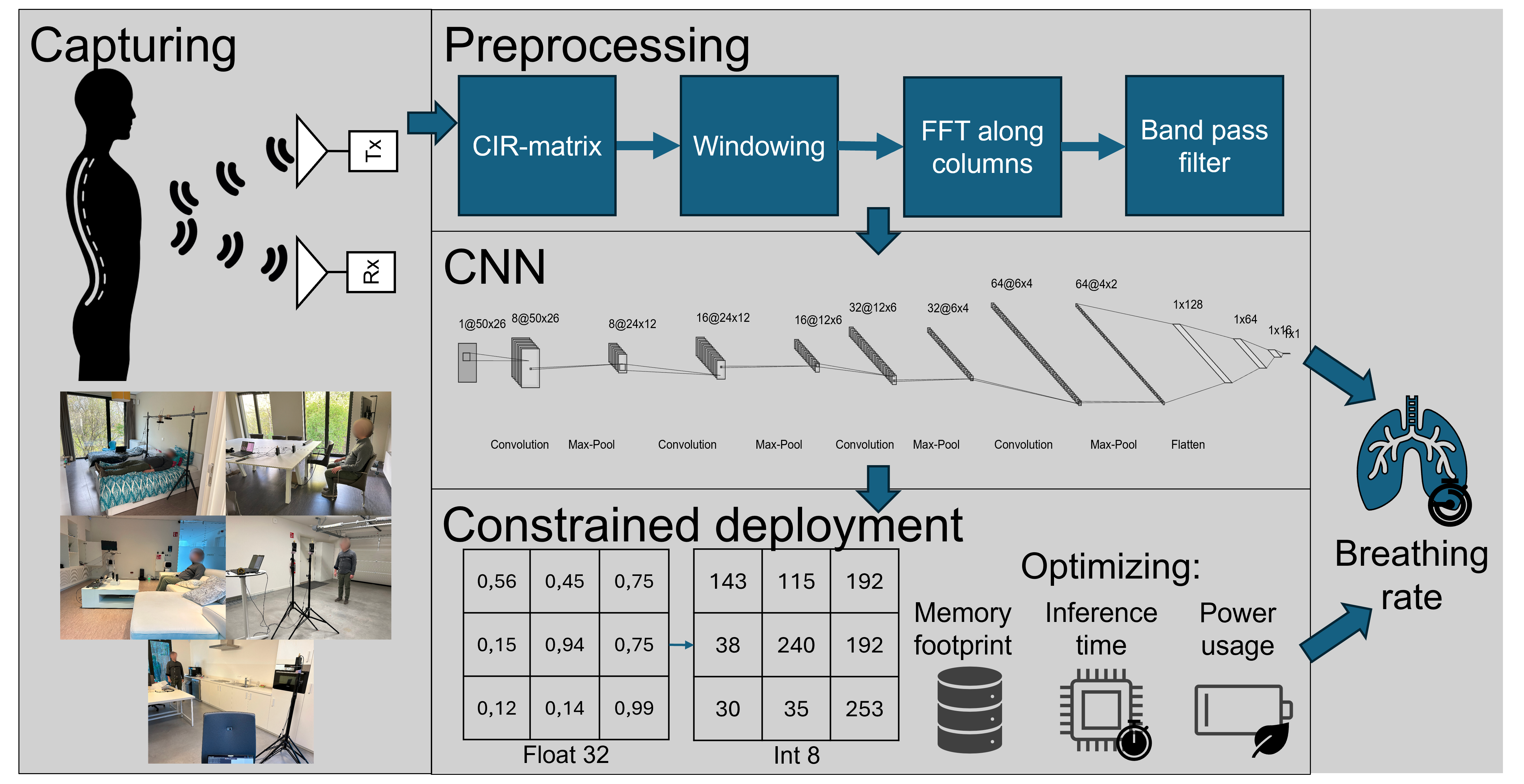}%
\end{wrapfigure}%
\begin{abstract}
Respiratory diseases account for a significant portion of global mortality. Affordable and early detection is an effective way of addressing these ailments. To this end, a low-cost commercial off-the-shelf (COTS), IEEE 802.15.4z standard compliant impulse-radio ultra-wideband (IR-UWB) radar system is used to estimate human respiration rates. We propose a convolutional neural network (CNN) \added[id=rev1]{specifically adapted} to predict breathing rates from ultra-wideband (UWB) channel impulse response (CIR) data, and compare its performance with both other rule-based algorithms \added[id=rev1]{and model-based solutions}. The study uses a diverse dataset, incorporating various real-life environments to evaluate system robustness. \added[id=rev1]{To facilitate future research, this dataset will be released as open source.} Results show that the CNN achieves a mean absolute error (MAE) of 1.73 breaths per minute (BPM) in unseen situations, significantly outperforming rule-based methods (3.40 BPM). By incorporating calibration data from other individuals in the unseen situations, the error is further reduced to 0.84 BPM. In addition, this work evaluates the feasibility of running the pipeline on a low-cost embedded device. Applying 8-bit quantization to both the weights and input/output tensors, reduces memory requirements by 67\% and inference time by 62\% with only a 3\% increase in MAE. As a result, we show it is feasible to deploy the algorithm on an nRF52840 system-on-chip (SoC) requiring only 46 KB of memory and operating with an inference time of only 199 ms. \added[id=rev1]{Once deployed, an analytical energy model estimates that the system, while continuously monitoring the room, can operate for up to 268.2 days without recharging when powered by a 20 000 mAh battery pack.} For breathing monitoring in bed, the sampling rate can be lowered, extending battery life to 313.8 days, making the solution highly efficient for real-world, low-cost deployments. 
\end{abstract}

\end{minipage}}}

\maketitle

\section{Introduction}
\label{sec:introduction}

\added[id=rev2]{According to Eurostat, respiratory diseases accounted for 6.1\% of all deaths in the EU in 2021 \cite{eurostat2023}. This statistic highlights the significant impact of respiratory issues on public health. Early detection effectively addresses these ailments \cite{bousquet_impact_2013, kostikas_clinical_2020}, as timely identification significantly improves recovery prospects and enables more effective symptom management. Proper monitoring of vital signs plays a significant role in facilitating this process, specifically \ac{br} is a key metric for this application \cite{kayser_respiratory_2023}.} A variety of devices already exist to monitor \ac{br} reliably. One example is a nasal pressure transducer, which measures airflow through the nostrils to track inhalation and exhalation phases. Another is the respiratory effort belt, worn around the chest or abdomen to monitor chest wall movement. While effective and widely used, these devices require direct skin contact, making them unsuitable for patient scenarios like burn victims, where attaching sensors could cause further harm. Additionally, constant monitoring throughout the day using contact-based solutions is inconvenient for individuals. A remote monitoring system could overcome these challenges, enabling respiratory data collection without causing discomfort.

\added[id=rev2]{Such systems are particularly valuable for long-term health tracking, including elderly care and sleep quality assessment, the latter of which serves as a vital proxy for physiological stress and alertness \cite{choshenhillel_acute_2021}. By eliminating the need for wearable hardware, \ac{br} can be monitored over extended durations, shifting the focus from occasional snapshots to the analysis of long-term physiological trends.}

\added[id=rev2]{This study utilizes a non-invasive \ac{ir-uwb} radar system to measure respiratory rates by capturing chest wall reflections. The system does not require on-body sensors and is installed in front of the person. This makes it well-suited for monitoring patients in bed, tracking sedentary users in an office or domestic environment, and observing drivers or passengers during stationary periods.}

For non-invasive breathing monitoring, multiple radar technologies are available, including mmWave, \ac{fmcw} radar or \ac{uwb} radar. In this study, \ac{uwb} radio technology was chosen for its use of large bandwidth, allowing the transmission of very short radio pulses lasting nanoseconds \cite{cheraghinia_comprehensive_2025}. These very short \ac{uwb} pulses and their reflections can be timestamped with high accuracy, allowing the detection of small chest movements. These precise distance measurements of chest movements are crucial for detecting breathing signals. Compared to alternative radar technologies, \ac{uwb} also requires lower power consumption, is more affordable and generates minimal interference with other radio signals. \added[id=rev2]{Moreover, \ac{uwb} radios can also be used for other purposes, such as communication or localization, making it a suitable technology for multi-functional systems that provide integrated sensing, localization and communication \cite{van2024wip, coppens_uwb_2026}.} Analyzing data from the radar hardware involves selecting from various processing techniques. While most studies on \ac{br} determination rely on signal processing methods, data-driven based approaches have shown promising results, particularly in challenging conditions \cite{Li_2024, Mu_2024}.




The key contributions of this work are the following:
\begin{itemize}
    \item \added[id=rev1]{A \ac{cnn}-based processing pipeline is proposed for breathing determination that shows significant accuracy improvement as compared to rule-based, and model-based processing techniques.}

    \item \added[id=rev1]{The proposed solution is evaluated with an improved strategy considering multiple environments, subjects, and orientations, providing a more thorough assessment of generalization to unseen conditions.}

    \item \added[id=rev1]{The algorithmic complexity of each processing step is analyzed, providing insights into scalability and computational feasibility on resource-constrained devices.} 

    \item \added[id=rev1]{The effects of model quantization on accuracy, memory usage, execution time, and energy consumption are modeled and evaluated, highlighting practical trade-offs for embedded deployment.}

    \item \added[id=rev1]{An analytical energy model is developed, covering the radar hardware, signal-processing pipeline and the model inference stage, enabling accurate predictions of execution time, memory footprint, and energy consumption.}

    \item An open source dataset is provided consisting of \ac{uwb} breathing data and annotated ground truth measurements captured using a low-cost \ac{cots} \ac{ir-uwb} device.\footnote{https://gitlab.ilabt.imec.be/datasets/uwb-vital-sign-monitoring}
    
\end{itemize}

The remainder of the paper is organized as follows. First, \Cref{sec:related_work} provides an overview of the current state of the art. Next, \Cref{sec:system_model} provides a mathematical model of the system. \Cref{sec:measurement_campain} describes how the dataset was captured. \Cref{sec:methodology} describes the algorithms developed to extract \ac{br} from the \ac{cir} data. \Cref{sec:results_and_analysis} analyses the obtained results. Afterwards, \Cref{sec:embedded_deployment} describes how the solution can be optimized and deployed on embedded hardware and analyses the reduced footprint, inference time, and power consumption. Finally, \Cref{sec:conclusion} concludes the paper and describes future work.

\section{Related work}
\label{sec:related_work}

\begin{table*}[]
\centering
\begin{tabular}{cccc c | cc | ccccccc | c}
\multicolumn{5}{c}{}                                                                       & \multicolumn{2}{c}{Low-cost}                                      &  \multicolumn{7}{c}{Real life considerations} \\\hline
                           & Year          & Device          & \makecell{Bandwidth\\(GHz)} & \makecell{IEEE\\comp-\\liant} & \makecell[c]{COTS}   & \makecell[c]{Embedded\\impl.}    & \makecell{Dist-\\ance}  & \makecell{Rotat-\\ion}  & \makecell{Environ-\\ment} & \makecell{Post-\\ure} & \makecell{Multiple\\persons\\in dataset} & \makecell{Multi-\\person}& \makecell{\added[id=rev2]{Non}\\\added[id=rev2]{static}}   & \makecell{\added[id=rev1]{AI-}\\ \added[id=rev1]{based}}   \\\hline\\
\cite{Leem_2017}           & 2017          & NVA6201         & 2.3                &  \xmark          &  \xmark                          &  \xmark                            &  \xmark  &  \xmark  &  \xmark  &  \xmark  & \cmark &  \xmark  & \added[id=rev2]{\cmark}  & \added[id=rev1]{\xmark} \\      
\cite{Duan_2019}           & 2019          & PulsOn 440      & 4.3                &  \xmark          &  \xmark                          &  \xmark                            & \cmark &  \xmark  &  \xmark  &  \xmark  & \cmark &  \xmark  &  \added[id=rev2]{\xmark}   & \added[id=rev1]{\xmark} \\      
\cite{Zhang_2020}          & 2020          & X4              & NA                 &  \xmark          &  \xmark                          &  \xmark                            &  \xmark  & \cmark &  \xmark  &  \xmark  & \cmark &  \xmark  &  \added[id=rev2]{\xmark}   & \added[id=rev1]{\xmark} \\      
\cite{Husaini_2022}        & 2022          & X4              & NA                 &  \xmark          &  \xmark                          &  \xmark                            & \cmark & \cmark &  \xmark  &  \xmark  &  \xmark  &  \xmark  &  \added[id=rev2]{\xmark}   & \added[id=rev1]{\xmark} \\      
\cite{Lopes_2022}          & 2022          & X4M03           & 1.4                &  \xmark          &  \xmark                          &  \xmark                            & \cmark &  \xmark  &  \xmark  &  \xmark  & \cmark &  \xmark  &  \added[id=rev2]{\xmark}   & \added[id=rev1]{\xmark} \\      
\cite{Farsaei_2023}        & 2023          & Proprietary     & 0.5                &  \cmark          &  \xmark                          &  \xmark                            &  \xmark  &  \xmark  &  \xmark  &  \xmark  & \cmark & \cmark &  \added[id=rev2]{\xmark}   & \added[id=rev1]{\xmark} \\      
\cite{Wang_2020}           & 2020          & X4              & 1.5                &  \xmark          &  \xmark                          &  \xmark                            & \cmark & \cmark & \cmark &  \xmark  &  \xmark  &  \xmark  &  \added[id=rev2]{\xmark}   & \added[id=rev1]{\xmark} \\      
\cite{Li_2024}             & 2023          & NVA6100         & NA                 &  \xmark          &  \xmark                          &  \xmark                            &  \xmark  &  \xmark  & \cmark & \cmark & \cmark &  \xmark  &  \added[id=rev2]{\xmark}   & \added[id=rev1]{\xmark} \\      
\cite{Mu_2024}             & 2024          & X4M03           & 1.4                &  \xmark          &  \xmark                          &  \xmark                            & \cmark & \cmark &  \xmark  &  \xmark  & \cmark &  \xmark  &  \added[id=rev2]{\xmark}   & \added[id=rev1]{\xmark} \\      
\cite{Wang_2024}           & 2024          & X4M05           & 1.4                &  \xmark          &  \xmark                          &  \xmark                            & \cmark &  \xmark  & \cmark &  \xmark  & \cmark &  \xmark  & \added[id=rev2]{\cmark}  & \added[id=rev1]{\xmark} \\      
\cite{Kim_2024}            & 2024          & X4M06           & NA                 &  \xmark          &  \xmark                          &  \xmark                            & \cmark & \cmark &  \xmark  & \cmark &  \xmark  &  \xmark  & \added[id=rev2]{\cmark} & \added[id=rev1]{\xmark}  \\      
\cite{Dang_2022}           & 2022          & X4m200          & 1.4                &  \xmark          &  \xmark                          &  \xmark                            & \cmark &  \xmark  & \cmark &  \xmark  & \cmark & \cmark &  \added[id=rev2]{\xmark}  & \added[id=rev1]{\xmark}  \\      
\cite{Khan_2017}           & 2017          & NVA6201         & 2.3                &  \xmark          &  \xmark                          &  \xmark                            & \cmark & \cmark &  \xmark  &  \xmark  & \cmark &  \xmark  & \added[id=rev2]{\cmark}  & \added[id=rev1]{\xmark} \\      
\cite{Zhou_2023}           & 2023          & X4M03           & 1.4                &  \xmark          &  \xmark                          &  \xmark                            & \cmark & \cmark &  \xmark  &  \xmark  & \cmark & \cmark & \added[id=rev2]{\cmark}  & \added[id=rev1]{\cmark} \\      
\cite{Numan_2023}          & 2023          & NCJ29D5         & 0.5                & \cmark         & \cmark                         &  \xmark                              &  \cmark  &  \xmark  &  \xmark  &  \xmark  & \cmark &  \xmark  &  \added[id=rev2]{\xmark}   & \added[id=rev1]{\xmark} \\      
\multicolumn{2}{c}{This paper}             & DW3000         & 0.5                & \cmark         & \cmark                         & \cmark                               & \cmark & \cmark & \cmark & \cmark & \cmark &  \xmark  &  \added[id=rev2]{\xmark}      & \added[id=rev1]{\cmark}      
\end{tabular}
\caption{Overview of related work, categorized by hardware type and application context. This work is the first to provide a truly low-cost, IEEE 802.15.4z-compliant, off-the-shelf available solution that can run on embedded hardware while coping with a variety of real-life environments and conditions. \added[id=rev1]{Indicated by the final column, using an \ac{ai}-approach in this context has not been explored extensively.}}
\label{tab:related_work}
\end{table*}

\Cref{tab:related_work} provides an overview of related scientific papers. As seen in the table, most studies employ radar systems designed by Novelda or Pulson devices, which consistently use high bandwidths of 1.4~GHz or higher. A higher bandwidth enhances the precision of the \ac{cir}s, aiding in capturing subtle breathing patterns. However, such systems require more powerful downstream processing, reducing the cost-effectiveness of the system. These high-bandwidth devices are also not compliant with the IEEE 802.15.4z standard, lacking standardized bandwidth and center frequency configurations. Additionally, they are significantly more expensive, costing at least 10 times more than commercial standard-compliant transceivers such as Qorvo’s DW3000, which is priced under \euro 10. Finally, most of these Novelda and Pulson devices are no longer commercially available. 

Furthermore, as discussed in \Cref{sec:introduction}, capturing breathing movements is challenging since individuals are not always stationary or aligned with radar devices. Validation on diverse real-life conditions is essential to claim robustness. Columns 7–13 of Table I summarize these considerations across the reviewed papers, covering tested distances and rotation angles, variation across environments and postures, the presence of multiple subjects in the dataset, multi-person scenarios (multiple subjects measured simultaneously), and subject motion during measurements.


When considering the solutions that use IEEE 802.15.4z-compliant hardware, the following work is most relevant. Farsaei et al. \cite{Farsaei_2023} utilize a 500 MHz standard-compliant radar system, but their hardware is proprietary and inaccessible, disqualifying it as \ac{cots}. Only 2 real-life considerations are tested using a peak-finding algorithm. In general, validation on \ac{br} determination is limited, as only 2 \ac{cir}-windows are considered in their validation set. Numan et al. \cite{Numan_2023} stand out for using the NCJ29D5 radar from NXP. This is a \ac{cots} device costing less than \euro 10. The paper only evaluates generalizability to new individuals and over multiple distances. This work focuses more on presence detection and classifying low, normal, and high \ac{br}s rather than predicting the exact respiratory rates. \ac{br} determination is done with a peak detection-based solution, and its validation is restricted to 9, 10-second \ac{cir} windows.

\added[id=rev1]{The final column of \Cref{tab:related_work} shows whether a solution uses \ac{ai}. All but one prior work utilizes a model-based \ac{br} estimator. This exception is Zhou et al. \cite{Zhou_2023}, which addresses a comparable number of challenges compared to this paper but lacks evaluations in multiple environments or with subjects in different postures. However, it does excel in analyzing non-static scenarios and multi-person setups. Their approach uses an Attention-Based \ac{cnn}-\ac{lstm} model to filter out motion. Afterwards, the filtered data is analyzed with a rule-based \ac{br} estimator. In contrast to our work the authors do not analyze the computational costs or provide energy models to evaluate the suitability for embedded implementations. The lack of environmental testing also raises major concerns about generalizability, a key hurdle in \ac{ai}-driven solutions.}

\added[id=rev1]{The work described in this paper stands out from all works in \Cref{tab:related_work} by using \ac{cots} IEEE 802.15.4z-compliant hardware, validating the system with multiple real-life considerations in mind and also analyzing the processing requirements of the solution. Compared to prior work, this study is the first to focus on analyzing the computational and energy requirement aspects required for embedded hardware compatibility.}

\section{System model}
\label{sec:system_model}

This study employs a radar-based system with a transmitter emitting signals that reflect off objects or people in the room. These reflections will be influenced by respiratory movements of the person in front of the radar and are subsequently captured by the receiver. The varying distance between the radar and the chest of the target is then used to extract the \ac{br} \cite{venkatesh_2005}. \Cref{form:model_breath} shows mathematically how this distance is modelled.

\begin{equation}
\label{form:model_breath}
d(t) = d_0 + m(t) = d_0 + m_b sin(2 \pi f_bt)
\end{equation}

\noindent where $f_b$ and $m_b$ are the breathing frequency and amplitude, respectively. The $d_0$ term represents the distance from the target to the radar. This paper only considers static targets, so in our work this term can be considered a constant. To extract respiratory information, \ac{uwb} radio pulse signals are transmitted by a transmitter at regular intervals with a time interval in the order of milliseconds. These transmissions are also called slow-time samples. For each slow-time transmission, multiple reflections are received by a receiver, which are sampled as complex \ac{iq} values with a resolution in the order of nanoseconds. Because of this higher sampling rate, these are also called fast-time samples. A collection of fast-time samples that correspond to 1 slow-time index forms a \ac{cir}. A collection of slow-time indices can then be represented as a matrix with the notation shown in \Cref{form:cir_matrix}.

 \begin{equation}
 \label{form:cir_matrix}
     R = {r[m, n]}, 1 \leq m \leq N_s, 1\leq n \leq N_f
 \end{equation}

\noindent where each row corresponds to a \ac{cir}, $N_s$ is the number of slow-time samples and $N_f$ is the number of fast-time samples per \ac{cir}. Another way to interpret these indices is to consider the slow-time axis as discrete time instants and the fast-time indices as distances from the radar to the reflecting surfaces. Each transmitted signal represents a time step. Within this step, captured fast-time samples correspond to reflections originating from increasingly further distances. Consequently, taking more fast-time samples at a single slow-time index provides reflection data from more distant surfaces. Reflections originate not only from the target but also from static objects in the room. Assuming these objects are static, the received signal can be expressed as \Cref{form:signal_clutter}.

\begin{equation}
\label{form:signal_clutter}
r(t, \tau ) = \sum_{i=0}^{} A_i p(\tau - \tau_i) + A_p p(\tau - \tau_{d(t)})
\end{equation}

\noindent where $t$ and $\tau$ are slow- and fast-times indices, respectively. The first term sums the static background reflections ($A_i$), also called clutter. The second term represents reflections from the target ($A_p$), where the time-dependent delay $\tau_d(t)$ varies with breathing. \Cref{fig:preprocessing_pipeline}a illustrates the amplitude of a single \ac{cir}. The \ac{fp} corresponds to where the amplitude of the captured \ac{iq} values rises above the noise floor. In \Cref{fig:preprocessing_pipeline}a, this occurs at index 0. 

\section{Measurement campaign and dataset description}
\label{sec:measurement_campain}

For evaluating the different algorithms, data was collected using the Qorvo DW3000, specifically the QM33120WDK1 \ac{dk}, acting as a representative low-cost, \ac{cots} hardware device. The setup includes separate transmitter and receiver devices, each equipped with a Nordic nRF52840 \ac{dk} and a DW3000 transceiver shield. The transmitter uses an omnidirectional antenna, while the receiver employs a directional antenna to focus on reflections coming from the desired direction. The system operates at a center frequency of 6.489~GHz with a bandwidth of 500~MHz. The slow-time sampling frequency is 77.5~Hz, and the receiver captures \ac{iq} samples at 1~ns intervals. Each fast-time sample represents 30~cm in distance, with each \ac{cir} containing 41 fast-time samples and the first peak always aligned at index 3. Samples before the first peak can be considered noise. The maximum distance the system can reach is equal to: $((41-4)*30~cm)/2=555~cm$. A Plux respiratory effort belt \cite{plux_respiration_2020} serves as ground truth, measuring the accurate \ac{br} through belt stretch.

\begin{figure}[!t]
\centerline{\includegraphics[width=1.0\columnwidth]{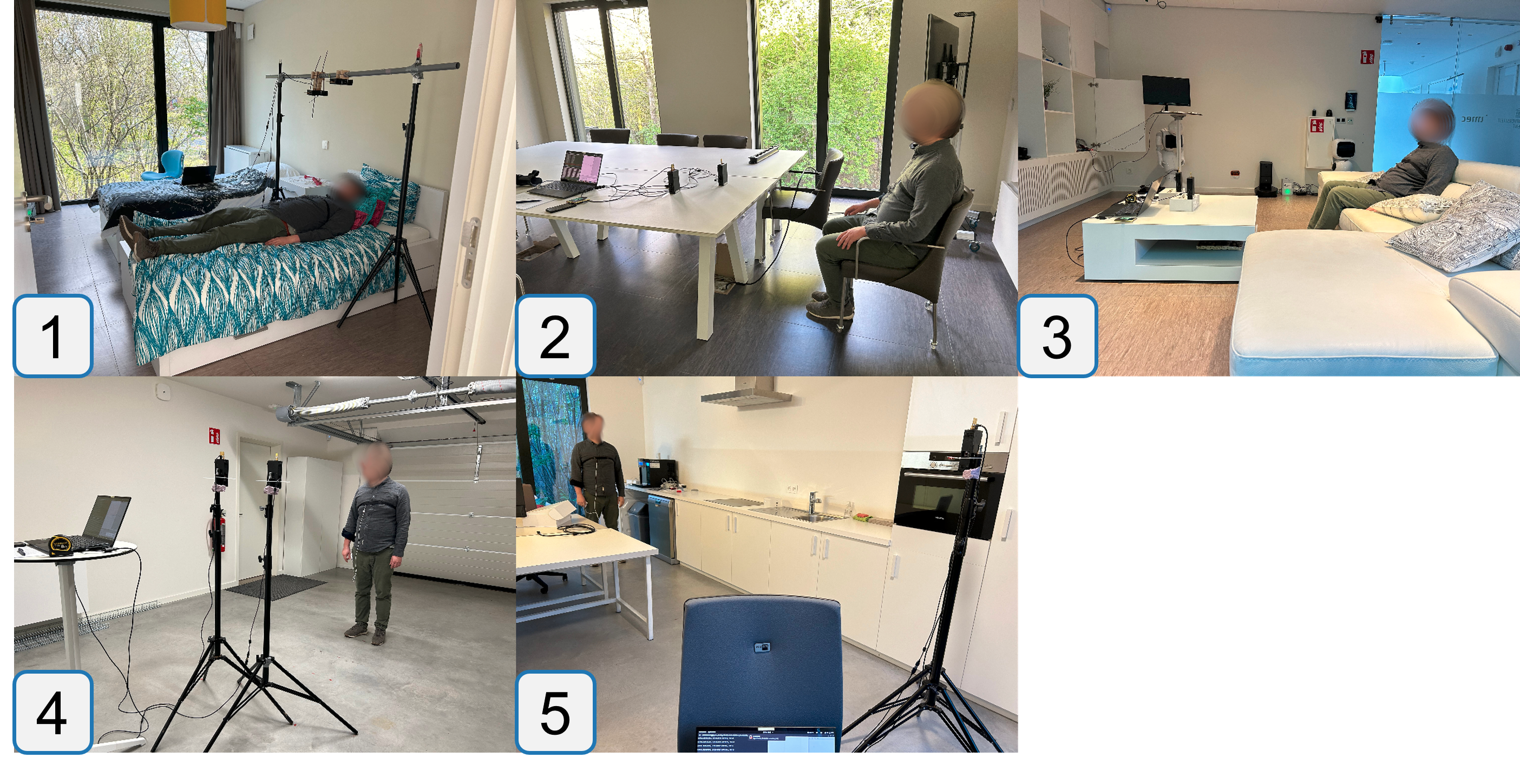}}
\caption{Samples were collected across various environments with 16 participants. Environment 1 had subjects lying down. Environments 2 and 3 had them sitting. In the final 2 environments, subjects stood at multiple distances and angles from the radar. Defining a setup as a distinct combination of environment, distance, angle, and posture resulted in 15 different setups.}
\label{fig:images_setup_homelab}
\end{figure}

The dataset was collected in the Homelab in Zwijnaarde \cite{idlab_idlab_2024}, a simulated home environment developed by Ghent University and imec for testing \ac{iot}, smart home, and healthcare applications. The lab features various residential rooms, enabling diverse testing conditions. \Cref{fig:images_setup_homelab} illustrates the environments used in this project.
To avoid saturation at the receiver side, the transmitter and receiver were placed 30 cm apart.
In the first environment, targets lay in bed with the \ac{uwb} hardware positioned 100~cm above their chest. In environments 2 and 3, participants sat on a chair at distances of 120~cm and 180~cm from the radar. Environment 4 involved standing individuals at varying distances (60, 100, 150, 200, and 400 cm) from the radar, with additional measurements at a 45-degree rotation angle for distances 60~cm and 150~cm. In the final environment, participants stood at the same distances without angle variations. For environments 4 and 5, both the transmitter and receiver were placed at a height of 156~cm. In total, 16 different people participated in the data campaign, resulting in a total of 240 different person-setup pairs\footnote{Informed consent was obtained from all participants prior to the start of the measurement campaign.}. In environments 1, 2 and 3, measurements of 2 minutes were taken from each person-setup pair. In environments 4 and 5, only 1 minute \ac{cir} windows were captured. Consequently, approximately 288~min of data was obtained.

To ensure variability in \ac{br}s, subjects followed a randomized breathing pattern guided by an app, producing rates between 6 and 20~\ac{bpm}. \added[id=rev2]{This range was selected to align with established physiological benchmarks. Porter et al. define the nominal resting respiratory rate for adults as 12–20~\ac{bpm} \cite{porter_abnormal_2025}. By extending the experimental range down to 6~\ac{bpm}, the study incorporates the detection of bradypnea (abnormally slow breathing), a clinical indicator correlated with specific stress responses and sleep-disordered breathing. High-frequency respiratory rates exceeding 20~\ac{bpm} (tachypnea) were excluded from this campaign, as such rates typically occur during physical exertion or acute clinical distress. Given that this work focuses on a non-intrusive baseline for resting-state and long-term trend monitoring, these higher frequencies were deemed outside the current scope.}

\section{Methodology}
\label{sec:methodology}

This section outlines the proposed processing pipeline, beginning with preprocessing steps and followed by different prediction algorithms. The prediction approaches are categorized into 2 types: rule-based peak-finding methods that serve as reference and the proposed model-based machine learning method.

\subsection{Preprocessing}
\label{sec:preprocessing}

\begin{figure*}[!t]
\centerline{\includegraphics[width=\textwidth]{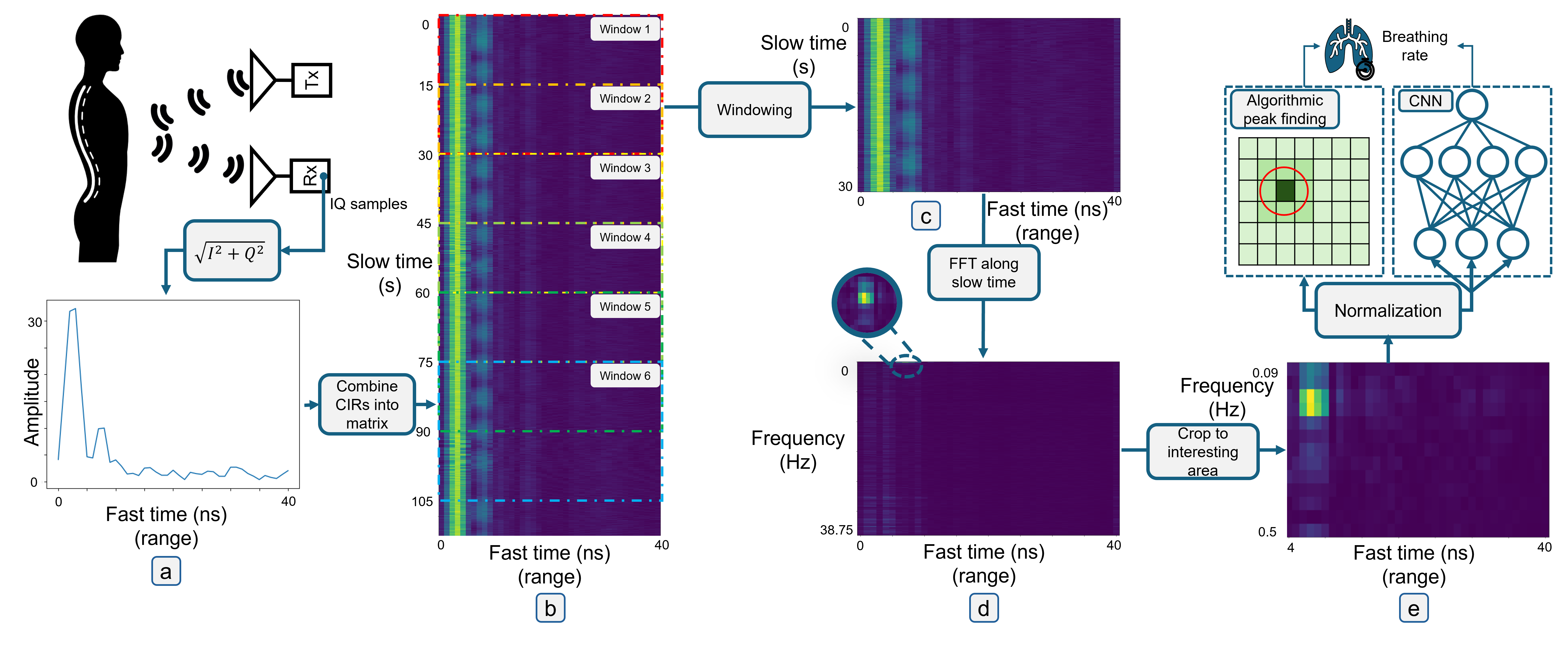}}
\caption{Visualization of processing pipeline. \ac{iq} samples coming from the receiver will serve as input for the pipeline. Each sub-representation of the data is described in Section \ref{sec:methodology} and is indicated with a letter at the bottom of the subfigure.}
\label{fig:preprocessing_pipeline}
\end{figure*}

The preprocessing pipeline is shown in Figs. \ref{fig:preprocessing_pipeline}a to \ref{fig:preprocessing_pipeline}e. This pipeline processes the samples captured as \ac{iq} samples. For each slow-time sample, 41 fast-time samples are taken, forming a single \ac{cir}. Once collected, the amplitude of these samples is calculated. The resulting amplitude versus fast-time is shown in \Cref{fig:preprocessing_pipeline}a. 
\added[id=rev2]{While phase information is available, it is discarded because phase stability across unsynchronized, low-cost UWB hardware is often unreliable. Furthermore, the high sensitivity of phase data makes it more vulnerable to environmental noise and multipath effects in complex settings.}
These \ac{cir}s are then combined into a matrix, of which a heatmap representation is shown in \Cref{fig:preprocessing_pipeline}b. Each row in the matrix corresponds to a \ac{cir} (the ``slow time"'), and each column captures variations across time at a certain distance (the ``fast time''). The maximum value of the slow-time index of the matrix is 120 seconds, representing all \ac{cir} samples collected from a single person in a specific environment and setup. 
To calculate the breathing rate over a shorter duration than the full experiment, a 30-second windowing operation is applied to the \ac{cir} matrix with a stride of 15 seconds. This window size was empirically determined to be an optimal balance between the dataset size required for model training and the accuracy of the \ac{fft} that will be applied later in the pipeline. Using an overlapping windowing approach results in more training samples, which is advantageous for training a \ac{nn}. \added[id=rev3]{The 15-second stride is used only during training to augment the dataset. At deployment, the update rate is configurable by adjusting the stride between successive windows.} Careful splitting between train and validation samples in our validation strategy will ensure no data leakage is introduced, which is further detailed in \Cref{sec:results_and_analysis}. An example of a window is shown in \Cref{fig:preprocessing_pipeline}c. In this representation, it is clear that a periodic signal is present around range bin 8. In the next step, an \ac{fft} is applied to each column, generating a frequency spectrum for each distance bin. The obtained representation is shown on \Cref{fig:preprocessing_pipeline}d, which is referred to as a range-frequency map. Peaks at the target breathing frequency are expected at the range index where a person is present. To focus on breathing rate-specific frequencies, signals below 0.09~Hz or above 0.5~Hz  should be filtered out. \added[id=rev1]{This band-pass operation also suppresses low-frequency components arising from static objects and slow-varying factors such as temperature or background changes.} When zooming in to this frequency range, peaks are indeed present in the example. Since the first peak index is present at index 3, the first 4 range bins can also be removed. Both the frequency filtering and range bin removal can be implemented with an image crop operation. The result of this operation is shown in \Cref{fig:preprocessing_pipeline}e, showing which movement frequencies are present at each distance. The final step involves applying Min-Max normalization to scale all values between 0 and 1. This ensures consistent data scales across all \ac{cir} windows, which will benefit \ac{nn} approaches. The min and max values are derived solely from the training data to prevent data leakage. The normalized \ac{cir} window then serves as input to the breathing detection algorithms, which in our case will be either rule-based techniques or neural networks.

\subsection{Peak finding techniques}
\label{sec:peak_finding}

As noted in \Cref{sec:related_work}, the most relevant low-cost or IEEE 802.15.4z-compliant studies are the papers by Numan et al. \cite{Numan_2023} and Farsaei et al. \cite{Farsaei_2023}, both of which use rule-based approaches. To enable comparison, this first category of prediction methods includes 3 different rule-based frequency peak-finding techniques. The first one closely resembles the approach by Farsaei et al., from which the others are 2 variations. All algorithms use the range-frequency maps as input, which can be mathematically represented as: $M = {m_{ij}}, 0 < i < n ; 0 < j < p$. Here, $m_{ij}$ is an element in the matrix, where $i$ represents the row index (frequency bin) and $j$ the column index (range bin). $n$ is the total number of frequency bins and $p$ is the number of fast-time indices. $f_i$ will be defined as the frequency associated with row $i$.

(i) The first approach identifies the breathing frequency by locating the row index $i_{max}$ of the maximum value in the matrix. Mathematically, this can be described as \Cref{form:max_peak}. Once $i_{max}$ is found, the corresponding frequency bin is: $f_{pred} = f_{i_{max}}$. From now on, this method is called the \textbf{highest peak algorithm}.

\begin{equation}
\label{form:max_peak}
   i_{max} = arg\max_{i}(\max_{j}\{m_{ij}\})
\end{equation}

(ii) In the second method, the distance-frequency map is first accumulated in 1 row with the operation described in \Cref{form:accumulation_operation}. After this, the highest frequency index can be determined with: $i_{max} = arg\max_{i}(a_i)$. This solution is the \textbf{accumulated highest peak algorithm}. The rationale is that scattering causes signals influenced by breathing movements to propagate to more distant surfaces. When these reflections are captured, the frequency component of the breathing signal is also present at these distances. By accumulating, it is hypothesized that the breathing frequency will become more pronounced.

\begin{equation}
\label{form:accumulation_operation}
   a_i = \sum_{j=1}^{p} m_{ij} , j = 1, 2, ..., p
\end{equation}

(iii) The last algorithm employs the same accumulated row resulting from \Cref{form:accumulation_operation}. This time, the predicted frequency is determined by the weighted average operation described in \Cref{form:weighted_avg}. Consequently, this solution is called the \textbf{accumulated weighted average algorithm}. Weighted averaging is also tested since this could alleviate the influence of large noise components.

\begin{equation}
\label{form:weighted_avg}
   f_{pred} = \sum_{i=1}^{n} \frac{a_i}{\sum_{k=1}^{n}{a_k}} f_i
\end{equation}

\subsection{Convolutional neural network based machine learning}
\label{sec:CNN_based_prediction}

Among existing studies, only Zhou et al. \cite{Zhou_2023} proposes an attention-based \ac{cnn}-\ac{lstm} solution evaluated on a similar amount of real-world considerations as our dataset. A more simple \ac{cnn} was chosen in this paper due to its significantly lower processing requirements, aligning better with the low-cost aspect of this paper. The model architecture and optimized parameters are detailed in \Cref{tab:CNN_architecture}. Various configurations of parameter sizes, kernel shapes, and layer counts were tested, with the best-performing design summarized in the table. A grid search was used to optimize normalization and regularization techniques, including batch normalization, dropout, and kernel regularization. The best-performing settings are also provided in \Cref{tab:CNN_architecture}. To ensure the robustness of the results, each model was trained 10 times over. The model that scored the median validation loss was selected. All subsequent reported results follow this training strategy.

\added[id=rev1]{
Although not previously applied to IR-UWB-based \ac{br} estimation, several resource-efficient model-based approaches exist, such as Random Forests and lightweight \ac{cnn}s like MobileNet. To provide constrained-\ac{ai} baselines, both methods were applied to our dataset. The Random Forest was implemented on flattened input features, while MobileNet used pretrained convolutional layers (ImageNet) with adapted fully connected layers for regression. To provide an additional \ac{cnn} baseline without an explicit emphasis on deployability, the  ResNet architecture was also evaluated with pretrained convolutional layers (ImageNet) and modified dense layers.}
\added[id=rev2]{Our proposed CNN (141~KB) is significantly more compact than the MobileNet (23.7~MB) and ResNet (89.6~MB) implementations. Efficiency is achieved through a tailored 36 × 13 input and a shallow four-layer structure. Aggressive pooling and minimal dense layers eliminate memory bottlenecks while effectively capturing frequency components across range bins. This compact design reduces overfitting risk and ensures compatibility with resource-limited hardware.}
\added[id=rev1]{More complex architectures (e.g., \ac{lstm} or transformers) were not considered, as they are incompatible with the embedded, resource-limited focus of this work.}

\begin{table}[]
\centering
\begin{tabular}{|p{0.28\columnwidth}|p{0.62\columnwidth}|}
\hline
Layer                     & Parameters\\
\hline
\ac{cir}\_window (input)       & size = (36, 13, 1)\\\hline
conv2d                    & \makecell[l]{\#filters = 8, kernel size = (10, 10),\\ activation = relu, padding = same,\\ kernel regularizer = L2}\\\hline
max\_pooling2d            & \makecell[l]{kernel size = (2, 2), strides = (2, 2)}\\\hline
conv2d\_1                 & \makecell[l]{\#filters = 16, kernel size = (8, 8),\\ activation = relu, padding = same,\\ kernel regularizer = L2}\\\hline
max\_pooling2d\_1         & \makecell[l]{kernel size = (2, 2), strides = (2, 2)}\\\hline
conv2d\_2                 & \makecell[l]{\#filters = 32, kernel size = (4, 4),\\ activation = relu, padding = same,\\ kernel regularizer = L2}\\\hline
max\_pooling2d\_2         & \makecell[l]{kernel size = (2, 2), strides = (2, 1)}\\\hline
conv2d\_3                 & \makecell[l]{\#filters = 64, kernel size = (2, 2),\\ activation = relu, padding = same,\\ kernel regularizer = L2}\\\hline
max\_pooling2d\_3         & \makecell[l]{kernel size = (2, 2), strides = (2, 1)}\\\hline
dense                     & \makecell[l]{dim = 64, activation = relu,\\ kernel regularizer = L2}\\\hline
dropout                   & dropout = 0.2\\\hline
dense\_1                  & \makecell[l]{dim = 16, activation = relu,\\ kernel regularizer = L2}\\\hline
dropout\_1                & dropout = 0.2\\\hline
dense\_2 (output)         & dim = 1\\
\hline
\end{tabular}
\caption{Architecture of the proposed CNN for breathing rate prediction.}
\label{tab:CNN_architecture}
\end{table}

\section{Results and analysis}
\label{sec:results_and_analysis}

This section evaluates the proposed \ac{cnn} and compares it to the representative rule-based solutions. This is done by applying the algorithms to the dataset described in \Cref{sec:measurement_campain}, comprising samples from 16 participants across 15 setups. Each of these participant-setup combinations forms a unique pair. For evaluating the generalizability of the solutions, specific pairs are excluded from the training data and used as validation samples. As a pair is always fully included or excluded from the training set, no data leakage will occur.

Three validation strategies are used to evaluate the generalizability of the model architecture. The leave-one-person-out cross-validation experiment assesses how well the proposed \ac{cnn} handles unseen individuals. The leave-one-situation-out cross-validation experiment evaluates its ability to adapt to new setups. Finally, the leave-one-person-situation-pair-out cross-validation strategy examines how the system performs with new combinations of individuals and setups. These strategies can also be interpreted as follows. If the system, trained on a certain set of person-situation pairs, is deployed to track an entirely new person, the leave-one-person-out strategy will estimate its performance. For deployment in a previously unseen setup, the leave-one-situation-out analysis indicates the expected performance. In the leave-one-person-situation-pair-out strategy, the scenario involves introducing a new setup, where measurements for some participants already in the dataset are used as calibration data. The prediction is that the calibration data will enhance the model's performance in the new setup.

\subsection{Generalization to new persons}
\label{sec:gen_to_new_people}

The first analysis evaluates the leave-one-person-out cross-validation strategy. In this approach, 1 participant is excluded from the training set in each fold and used for validation. This process is repeated for all 16 participants, producing 1 model per participant. The results are summarized in \Cref{tab:results_leave_one_person_out_cross_val}, showing the mean L1 errors and their standard deviation. For the highest peak method, mean L1 errors range from 3.10~\ac{bpm} to 5.58~\ac{bpm}. The accumulated distance-frequency map approach reduces mean L1 errors slightly, ranging from 2.75~\ac{bpm} to 5.32~\ac{bpm}. The weighted average algorithm shows L1 errors from 3.29~\ac{bpm} to 4.46~\ac{bpm}. Among the rule-based methods, selecting the highest peak from the accumulated frequency bins proves most effective. 

\begin{figure}[h]
\centerline{\includegraphics[width=\columnwidth]{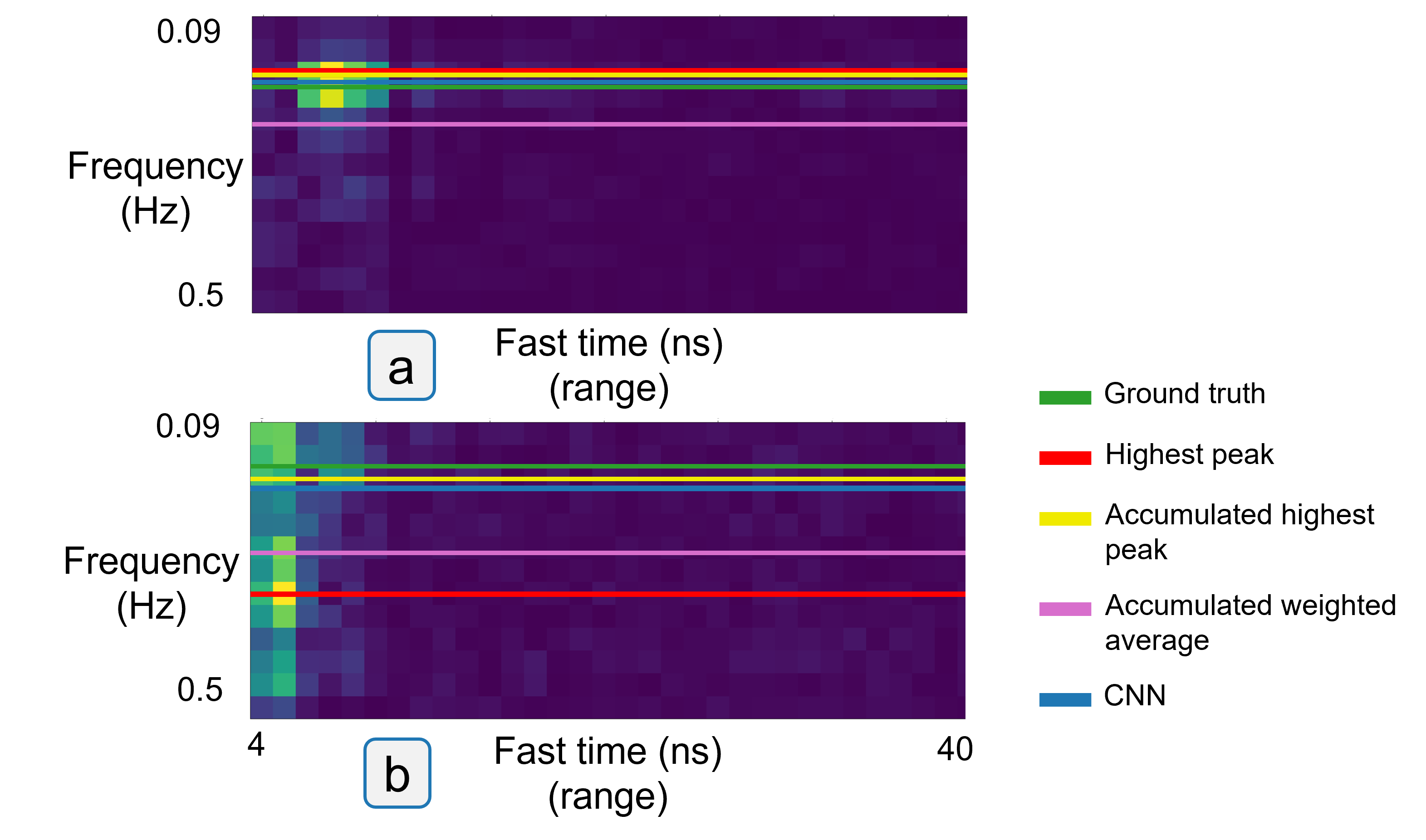}}
\caption{The breathing rate estimates of the rule-based algorithms and the proposed CNN are visualized on a range-frequency map. In a simple scenario with a visible peak, shown in (a), the predictions from all methods closely align. However, in noisier scenarios where the peak is less distinct, as shown in (b), the \ac{cnn} predictions are significantly closer to the ground truth.}
\label{fig:visualization_predictions_of_algorithms}
\end{figure}

\added[id=rev1]{The table also includes results from other \ac{ai}-based approaches previously mentioned in \Cref{sec:methodology}. The results in the rightmost column of \Cref{tab:results_leave_one_person_out_cross_val} show the performance of our proposed model. Overall, model-based solutions clearly outperform rule-based methods. \Cref{fig:visualization_predictions_of_algorithms} highlights conditions where the model-based solutions surpass traditional approaches. In `simple' conditions with little noise (\Cref{fig:visualization_predictions_of_algorithms}a), the breathing frequency peak is easily identified by most algorithms. In more challenging scenarios, however, noise components may dominate the breathing signal (\Cref{fig:visualization_predictions_of_algorithms}b). In these cases, the model-based solutions perform better by not only analyzing the highest peak, but also taking the higher harmonic components into account. Among all investigated solutions, our proposed \ac{cnn} achieves the best performance, with mean L1 errors between 0.97~\ac{bpm} and 2.51~\ac{bpm}, demonstrating a clear improvement over both rule-based methods and current \ac{ai} solutions. This clearly demonstrates that the architecture proposed in this work is well-optimized for effective performance on \ac{ir-uwb} data.}

When compared to prior work, Farsaei et al. \cite{Farsaei_2023} and Numan et al. \cite{Numan_2023} report mean L1 errors of $<$1~\ac{bpm} and 1.22$\pm$0.63~\ac{bpm}, respectively. In comparison, the rule-based approaches evaluated here show worse L1 scores. Our \ac{cnn} comes closer to this error but still results in a higher L1 error. This difference can be attributed to the inclusion of significantly more challenging scenarios and greater variability in \ac{br}s in our study. A fairer comparison is with the results from Zhou et al. \cite{Zhou_2023}, who report a mean L1 error of 2.3$\pm$1.9~\ac{bpm} across various challenging scenarios. Our proposed \ac{cnn} outperforms their results, with all validation folds in \Cref{tab:results_leave_one_person_out_cross_val} achieving lower mean L1 scores.

\added[id=rev1]{Collecting datasets in healthcare settings is resource-intensive and requires sufficient participant diversity to ensure statistical validity and generalizability. To assess how performance scales with sample size, we employed an extended leave-k-people-out cross-validation with multiple permutations of training–testing splits. Results show that L1 error consistently decreases as the number of training participants increases, though with diminishing returns: expanding from 1 to 5 participants reduced error by 0.262 BPM, from 5 to 10 by 0.261 BPM, and from 10 to 16 by 0.154 BPM. This indicates that while additional data may enhance performance, the incremental gains are expected to diminish quickly as more individuals are added.}

\begin{table*}[h]
\centering
\begin{tabular}{l|ll|ll|ll|ll|ll|ll|ll}
 & \multicolumn{6}{c}{Rule based}&  \multicolumn{2}{|c|}{AI based}&\multicolumn{6}{c}{Constrained AI based}\\
\hline
\multicolumn{1}{c|}{} &  \multicolumn{2}{c}{\makecell[c]{Highest\\peak}} & \multicolumn{2}{c}{\makecell[c]{Accumulated\\highest\\peak}} & \multicolumn{2}{c|}{\makecell[c]{Accumulated\\weighted\\average}}  &   \multicolumn{2}{c|}{\added[id=rev1]{ResNet50}}&\multicolumn{2}{c}{\added[id=rev1]{Random forest}}&\multicolumn{2}{c}{\added[id=rev1]{Mobilenet}} & \multicolumn{2}{c}{\makecell[c]{CNN\\\textbf{(Our solution)}}}\\

\hline

 Person & \makecell[c]{mean\\(\ac{bpm})} & \makecell[c]{std\\(\ac{bpm})} & \makecell[c]{mean\\(\ac{bpm})} & \makecell[c]{std\\(\ac{bpm})} & \makecell[c]{mean\\(\ac{bpm})} & \makecell[c]{std\\(\ac{bpm})} &  \makecell[c]{\added[id=rev1]{mean\\(\ac{bpm})}}   & \makecell[c]{\added[id=rev1]{std\\(\ac{bpm})}}  &\makecell[c]{\added[id=rev1]{mean\\(\ac{bpm})}}  &\makecell[c]{\added[id=rev1]{std\\(\ac{bpm})}}  &\makecell[c]{\added[id=rev1]{mean\\(\ac{bpm})}}   &\makecell[c]{\added[id=rev1]{std\\(\ac{bpm})}}   & \makecell[c]{mean\\(\ac{bpm})} &\makecell[c]{std\\(\ac{bpm})}  \\
 
\hline

       1  & 3.94               & 4.20              & 3.13                               & 2.79                              & 3.89                                 & 2.18                                &  2.04& 1.51&2.05 &1.30&2.03& 1.49& 1.65      &1.24      \\
       2  & 4.20               & 4.02              & 4.05                               & 3.85                              & 4.34                                 & 2.51                                &  2.59& 1.6&2.40 &1.43&2.55& 1.72& 2.08      &1.52      \\
       3  & 5.38               & 4.34              & \underline{\color{red}5.32}        & 4.89                              & 3.54                                 & 2.50                                &  \underline{\color{red}3.09}& 2.2&\underline{\color{red}2.82}&2.31&\underline{\color{red}2.77}& 2.1& \underline{\color{red}2.51}      &2.33      \\
       4  & 4.31               & 4.14              & 2.87                               & 3.32                              & 3.85                                 & 2.27                                &  2.47& 1.28&2.13 &1.41&2.47& 1.29& 1.51      &1.30      \\
       5  & 4.57               & 4.25              & 3.76                               & 3.97                              & 3.57                                 & 2.39                                &  2.7& 1.56&2.34 &1.36&2.59& 1.56& 1.59      &1.27      \\
       6  & 4.15               & 3.48              & 3.83                               & 3.54                              & 3.64                                 & 2.63                                &  2.74& 1.21&2.57 &1.32&2.69& 1.63& 2.06      &1.5       \\
       7  & 3.61               & 3.15              & 2.97                               & 3.18                              & 3.57                                 & 1.88                                &  \underline{\color{dark_green}1.97}& 1.19&\underline{\color{dark_green}1.68}&1.19&\underline{\color{dark_green}1.96}& 1.24& \underline{\color{dark_green}0.97}      &0.83      \\
       8  & 4.94               & 4.19              & 3.58                               & 3.70                              & 3.80                                 & 2.66                                &  2.88& 1.4&2.46 &1.48&2.87 & 1.37& 2.17      &1.57      \\
       9  & \underline{\color{red}5.58} & 4.67     & 4.10                               & 3.67                              & 3.99                                 & 2.56                                &  2.57& 1.4&2.47 &1.52&2.58 & 1.39& 2.05      &1.80      \\
      10  & 4.19               & 4.53              & \underline{\color{dark_green}2.75}      & 3.06                              & 4.26                                 & 2.32                                &  2.23& 1.38&1.88 &1.22&2.18 & 1.39& 1.35      &1.18      \\
      11  & \underline{\color{dark_green}3.10} & 3.40     & 3.19                               & 3.65                              & 4.25                                 & 2.34                                &  2.42& 1.6&2.16 &1.30&2.34& 1.75& 1.64      &1.48      \\
      12  & 3.67               & 3.14              & 3.33                               & 2.99                              & \underline{\color{red}4.46}                        & 2.19                                &  2.3& 1.64&2.12 &1.26&2.28& 1.3& 1.83      &1.34      \\
      13  & 4.19               & 4.49              & 2.98                               & 2.89                              & 4.10                                 & 2.27                                &  2.02& 1.32&1.94 &1.29&2.01& 1.51& 1.89      &1.29      \\
      14  & 3.29               & 3.26              & 2.80                                & 3.10                              & 3.98                                 & 2.21                                &  2.27& 1.44&2.20 &1.52&2.28& 1.42& 1.54      &1.38      \\
      15  & 3.52               & 3.78              & 2.86                               & 3.41                              & \underline{\color{dark_green}3.29}        & 2.42                                &  2.56& 1.45&1.99 &1.30&2.52& 1.52& 1.34      &1.11      \\
      16  & 3.61               & 3.74              & 3.01                               & 3.92                              & 4.22                                 & 2.07                                &  2.32& 1.96&2.24 &1.20&2.29& 1.68& 1.19      &0.97      \\
\hline
 \added[id=rev1]{Average}& 4.14& 3.92& 3.41& 3.50& 3.92& 2.34& 2.45& 1.51& 2.22& 1.40& 2.40& 1.52& 1.71&1.38\\

\end{tabular}
\caption{Mean and standard deviation of the L1 error obtained when predicting the breathing rate of unseen persons. The CNN performs better than rule-based \added[id=rev1]{and other \ac{ai}-based techniques}, with mean accuracies between 0.97 and 2.51 BPM and lower standard deviations.}
\label{tab:results_leave_one_person_out_cross_val}
\end{table*}

\subsection{Generalization to new situations}

The leave-one-setup-out cross-validation strategy evaluates generalizability to unseen setups. In each fold, 1 setup is entirely excluded for validation.
\Cref{fig:leave-one-situation-out_and_pair_out_cross} presents the results, where red box plots show the \ac{cnn} performance, and the green, yellow, and purple plots represent the highest peak, accumulated highest peak, and accumulated weighted average algorithms. The box plots display L1 errors across validation samples, with the x-axis indicating the setup parameters in the validation set. The results confirm that the \ac{cnn} generally outperforms rule-based methods. Only in setup 1 do the highest peak and accumulated highest peak algorithms slightly outperform the \ac{cnn}. However, this difference is negligible. 

\added[id=rev1]{Rule-based results in ``simple'' scenarios (setups 1 and 2) are comparable to prior work by Farsaei et al. \cite{Farsaei_2023} and Numan et al. \cite{Numan_2023}. However, in more challenging conditions not addressed by these prior studies, their performance degrades sharply, while the proposed \ac{cnn} maintains reasonable accuracy. Validation folds for environments 1, 2, and 3 involved entirely unseen surroundings. The model’s strong performance in these cases demonstrates its robustness to environmental variation, a common challenge for many \ac{ai}-based methods. Comparable evidence of generalizability is also absent in Zhou et al. \cite{Zhou_2023}.}

Finally, we consider a situation whereby a situation has been observed, but a new person is monitored for which the \ac{cnn} was not trained. This ``leave-one-person-setup-pair-out'' cross-validation strategy assesses performance when data from a specific person-setup pair is excluded.
The results are represented by the blue box plots in \Cref{fig:leave-one-situation-out_and_pair_out_cross}. These box plots show that when training includes data from other individuals in the same setup, the \ac{cnn} significantly improves its performance in fully unseen setups. With this strategy, only setup 10 has a Q3 above 2~\ac{bpm}, indicating robust performance in nearly all scenarios.

\subsection{Influence of situation parameters}

\begin{figure*}[!t]
\centerline{\includegraphics[width=1.0\textwidth,trim=4 4 4 4]{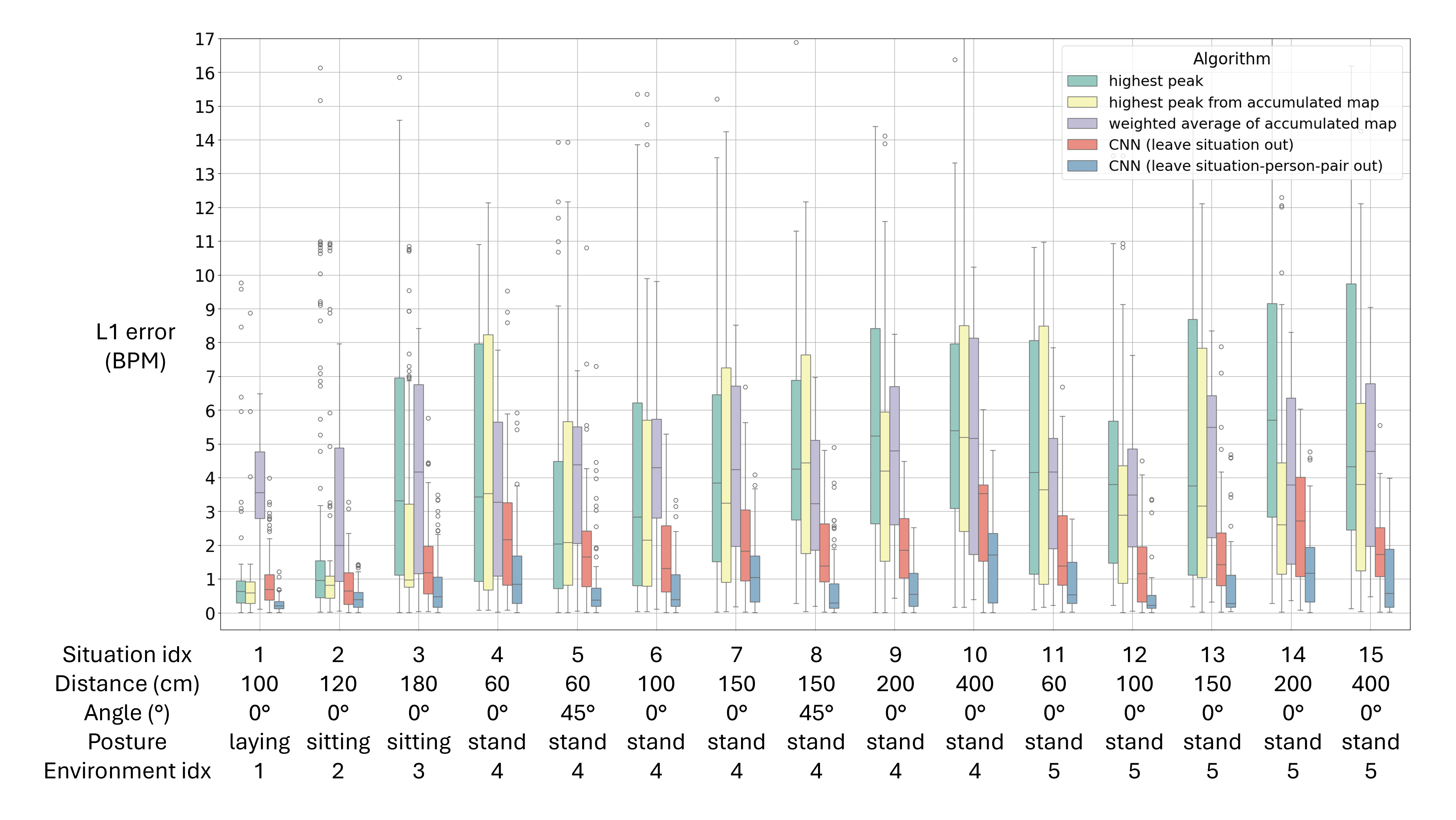}}
\caption{Breathing rate prediction errors in different situations (environment, posture, distance and body angle). The green, yellow and purple plots are the results when applying the rule-based methods described in \Cref{sec:peak_finding}. The red plots correspond to the L1 errors of the \ac{cnn} for fully unseen situations. The blue box plots correspond to the L1 errors of the \ac{cnn} for predicting the breathing rate for a new unseen person when other persons have already been observed in the specific situation (e.g. some form of pretraining on other individuals was possible). The \ac{cnn} achieves a \ac{mae} of 1.73 \ac{bpm} in unseen situations, significantly outperforming rule-based methods (3.40 BPM). When pretraining is possible, the \ac{mae} of the \ac{cnn} is further reduced to 0.84 BPM.}
\label{fig:leave-one-situation-out_and_pair_out_cross}
\end{figure*}

Based on the results in \Cref{fig:leave-one-situation-out_and_pair_out_cross}, several conclusions can be made concerning the influence of setup parameters on \ac{cnn} performance. In this analysis, the results from the leave-one-person-setup-pair-out cross-validation are considered. First, the posture of the target person significantly impacts performance. Analyzing individuals lying down is the easiest due to minimal \ac{rbm}. Sitting down introduces additional challenges due to increased reflection angles and slightly higher \ac{rbm}. Standing is the most difficult posture, consistently yielding poorer results at the same distances. The effect of distance is best represented in the scenarios where the individuals were standing. The accuracy generally decreases as the distance between the target and radar increases from 100 to 400 cm due to the decreased strength of the reflected signal. However, the \ac{cnn} still provides an acceptable mean accuracy of 1.29~\ac{bpm} at a 4-meter distance. Interestingly, there is a significant performance drop at a distance of 60 cm. This results from the fixed radar height of 156~cm (as noted in \Cref{sec:measurement_campain}) for environments 4 and 5. This height exceeds typical chest level, causing signals to arrive at steeper angles for shorter individuals, especially at close range. Additionally, at 60~cm, target reflections may overlap with first peak signals from the transmitter. For use cases where shorter ranges are required, the transmitter and receiver can be set up closer together. This would alleviate first peak overlap for shorter ranges but would require the transmission power to be lower to limit the saturation of the receiver. Finally, regarding the body angle, the \ac{cnn} provides an acceptable mean error of 0.86~\ac{bpm} even when the target person is positioned at a 45-degree angle towards the radar. This demonstrates that the \ac{cnn} can handle such scenarios effectively.

\subsection{Window Length Sensitivity Analysis}

\added[id=rev2]{The window duration $T$ determines the spectral resolution of the FFT preprocessing step ($\Delta f = 1/T$). While longer windows provide finer granularity, they increase temporal smoothing and increase computational complexity. Evaluating T between 20 and 40 s showed that \ac{mae} generally decreases as spectral resolution improves (from 1.84 BPM at 20 s to 1.62 BPM at 30 s and 1.57 BPM at 40 s). Despite the modest accuracy gain at larger window sizes, larger range-frequency maps increase input dimensionality, memory usage (from 141 KB at 30 s to 174.3 KB at 40 s), and inference time, while also extending settling time. Consequently, window length constitutes a design trade-off between estimation accuracy and responsiveness. A 30-second window was selected as a balanced operating point, but the optimal choice remains application-dependent.}

\section{Deploying on embedded hardware}
\label{sec:embedded_deployment}

Finally, this section thoroughly evaluates the feasibility of deploying the proposed solution on constrained hardware. First, the radar's slow-time sampling frequency is analyzed. The second subsection will give an assessment of the preprocessing pipeline's computational complexity. In the following sections, the resource usage of the \ac{cnn}-based solution is examined. The evaluation is conducted by deploying the model on the nRF52840 \ac{soc} to verify whether the model can run directly on the receiver device. In the final section, the power usage of the complete solution is described.

\subsection{Required sampling rate}
\label{sec:required_samp_rate}

\begin{figure}[!t]
\centering
\includegraphics[width=1.0\columnwidth]{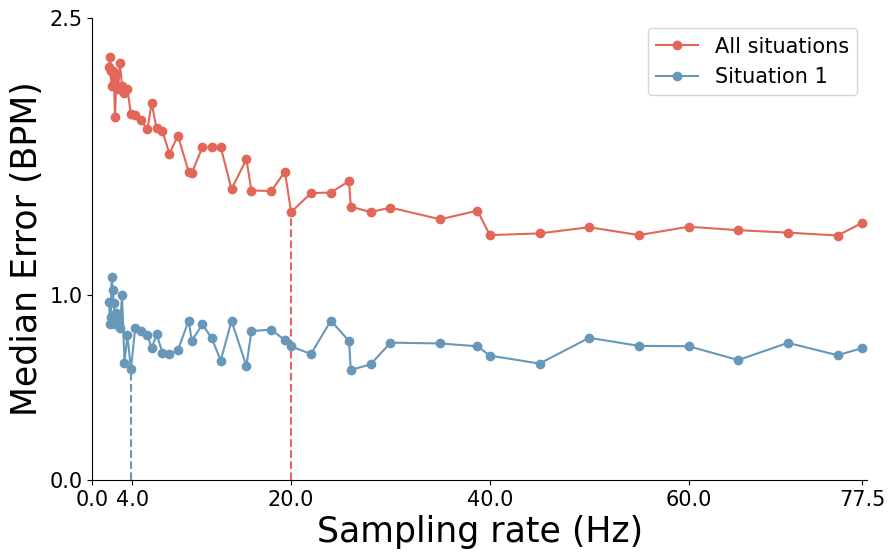}
\caption{Median L1 error for different sampling rates. Lowering the sampling rate negatively affects L1 error. For easy situations (situation 1), the sampling rate can be lowered to 4 Hz. When also considering more difficult situations, sampling rate should remain above 20 Hz.}
\label{fig:sub_sample}
\end{figure}

This section addresses the question of how often the \ac{uwb} transmitter should transmit a packet. The maximum sampling rate of our system is empirically determined to be 77.5~Hz\footnote{The maximum sampling rate is determined by the speed at which the \ac{cir} can be read out over the serial bus. Currently, 41 fast-time \ac{iq} samples are collected per \ac{cir}. Increasing the number of collected fast-time \ac{iq} samples will increase the maximum radar detection range but will decrease the maximum sampling rate at which \ac{cir} values can be read out.}. A higher sampling frequency requires sending more packets, which is not desirable for battery-powered embedded solutions. The red curve in \Cref{fig:sub_sample} illustrates the leave-one-setup-out cross-validation results on subsampled datasets. Limiting the sampling rate to 20 Hz has minimal impact on L1 error, but reducing it further significantly increases error. Given that human \ac{br}s peak at 0.5 Hz, Nyquist's theorem requires a minimal sampling rate of 1.0 Hz. The blue curve, representing the errors obtained when only validating on situation 1, shows that lower sampling rates have less impact in less complex conditions, with performance declining only at 4 Hz.


\subsection{Algorithmic complexity of preprocessing pipeline}

This section theoretically analyzes which parameters most profoundly impact the computation complexity of the preprocessing pipeline. First, the pipeline applies an \ac{fft} to each 30-second range bin. \added[id=rev1]{This \ac{fft} operation has a complexity of $O(n~\log~n)$, with $n$ the array length \cite{RAJABY2022103403}.} Since an \ac{fft} is performed for each fast-time index, indicated by $m$, the total complexity becomes $O(mn\log~n)$. Next, a cropping operation reduces the window size from $(n, m)$ to $(k, r)$, where $k$ is the number of frequency bins and $r$ is the number of range bins after cropping. This slicing operation has a complexity of $O(kr)$. Finally, a normalization operation is applied by using an element-wise subtraction and division operation. Both these operations have a complexity of $O(k~r)$. All previous steps can be summarized as $O(mn~\log~n + 3kr)$. Since $m$ and $r$ are radio configuration-dependent constants, this simplifies to $O(n\log~n + k)$. The dominant term, being $O(n\log~n)$, indicates an acceptable computational load suitable for constrained hardware.

\subsection{Memory usage of the \ac{cnn} model}

One consideration with the \ac{cnn}-based solution is its higher memory requirements compared to rule-based approaches. As the model size grows, the memory demand increases, limiting deployment on embedded devices. The proposed \ac{cnn} architecture occupies 141~KB. This is already an acceptable model size as the nRF52840 \ac{soc}, with 256~KB \ac{sram} and 1~MB flash memory available, can run this model. Further memory reduction can be achieved through quantization. This technique lowers the precision of the model parameters, converting the standard 32-bit floating-point values to lower-bit formats. In this case, the model parameters will be converted to 8-bit unsigned integers. In addition to quantizing the model parameters, the input and output tensors are also quantized, resulting in the entire network operating in 8-bit integer format. After this operation, the model size is reduced by 67~\% or to 46~KB, at the cost of only a 3.15~\% increase in \ac{mae} due to the mismatch between the float32 to int8 transformation operation, making it suitable for severely memory-constrained devices. 

\subsection{Execution time of the system}
\label{sec:time_complexity_of_model}

Next, we experimentally measure the execution time of the different approaches. Using the rule-based techniques results in an inference time of 0.139 ms. The inference time of the \ac{cnn} is higher at 523~ms. However, since 30-second windows are processed, the CNN's inference time is still significantly shorter than the considered time window. Reducing the inference time is still beneficial for embedded devices because this leaves more processing time for other tasks and allows the hardware to enter energy-saving states sooner. Quantization, as described earlier, not only reduces memory usage but also lowers inference time. When deploying the quantized model on the nRF52840 \ac{soc}, the inference time is reduced to  199~ms, which is a 62\% reduction. The nRF52840 \ac{soc} includes a \ac{fpu}, which helps handle floating-point operations. On microcontrollers without this support, the benefits of the quantized model would be even more pronounced.

\subsection{Energy consumption}
\label{sec:energy_consumption}

\subsubsection{Experimental measurement of the energy consumption}

\added[id=rev1]{
This section analyzes the energy consumption of the proposed solution. Practical measurements were obtained by deploying the complete processing pipeline on the nRF52840 \ac{dk} platform and monitoring current draw with a Keysight N6705B DC Power Analyzer at a supply voltage of 5~V (USB-powered). Reported currents represent averages, while power values are derived from precise area-under-the-curve calculations. These measurements were combined into a full-system energy model (\Cref{form:energy_model}), which accounts for platform ($P_{plat}$), transmitter ($P_{TX}$), and receiver ($P_{RX}$) consumption. \added[id=rev3]{For this analysis, preprocessing and model inference are assumed to occur once every 30 s, so they are included as $E_{preproc}$ and $E_{model}$, respectively. \Cref{form:energy_model} is structured so that readers can recompute the energy budget for any other update rate by replacing the 30-second interval in the $\lfloor t/30 \rfloor$ term with the desired stride.} As transmitter and receiver are separate devices, $P_{plat}$ is counted twice, with all processing executed on the receiver side.}

\added[id=rev1]{
Per-component measurements are summarized in \Cref{tab:power_con_per_comp}. The nRF52840 \ac{dk} consumes $1.84~mA$ at 5~V. Preprocessing is executed once every 30~s, drawing $2.03~mA$ for $0.110~s$. \ac{cnn} inference requires $4.57~mA$ for $0.523~s$, which is reduced to $4.29~mA$ and $0.199~s$ after quantization, yielding substantial savings. At a sampling rate of 20~Hz, which is shown in \Cref{sec:required_samp_rate} to provide robust operation under diverse conditions, the total energy consumption per 30~s is $E_{30sec} = 3.813~J$, corresponding to an average power of $0.127~W$. To put this in intuitive terms, a typical AAA cell stores roughly $5400~J$, so the system consumes one AAA's worth of energy every 11.8~h. When the sampling rate is reduced to 4~Hz for simpler scenarios, transmitter and receiver consumption decreases, lowering $E_{30sec}$ to $3.507~J$ (average power $0.117~W$). Preprocessing at this rate showed no measurable difference. Under this configuration, the system consumes one AAA's worth of energy every 12.8~h.
}





\added[id=rev1]{
\begin{equation}
\label{form:energy_model}
\begin{split}
    E_{\text{system}} &= \\
    &\underbrace{(P_{\text{plat}} + P_{\text{TX}})\cdot t}_{\text{transmitter}} +\\
    &\underbrace{(P_{\text{plat}} + P_{\text{RX}})\cdot t 
    + (E_{\text{preproc}} + E_{\text{model}}) \cdot \left\lfloor \tfrac{t}{30} \right\rfloor}_{\text{receiver}}
\end{split}
\end{equation}
}


\begin{table}
    \centering
    \begin{tabularx}{\columnwidth}{XXXX}\hline
 &  V (V)& P (W) & T (s)\\\hline
 \multicolumn{3}{c}{\textbf{Practical}}\\\hline
Platform&  5&  0.00923 & continuous \\ 
Preprocessing&  5&  0.01016 & 0.110 \\ 
Model&  5& 0.02287 & 0.523\\ 
Quantized~model&  5&  0.02146 & 0.199\\ 
 $TX_{20\_Hz}$&  5& 0.04474 & continuous \\ 
 $RX_{20\_Hz}$&  5& 0.06373 & continuous\\ 
 $TX_{4\_Hz}$&  5& 0.04357 & continuous\\ 
 $RX_{4\_Hz}$&  5& 0.05468 & continuous\\ \hline
  \multicolumn{3}{c}{\textbf{Theoretical}}\\\hline
 $TX_{20\_Hz}$&  3&0.000838 & continuous\\ 
 $RX_{20\_Hz}$&  3&0.001252 & continuous\\ 
 $TX_{4\_Hz}$&  3&0.000168 & continuous\\ 
 $RX_{4\_Hz}$&  3&0.000251 & continuous\\  
    \end{tabularx}
    \caption{
    \added[id=rev1]{Voltage and power consumption of components in the processing pipeline. Preprocessing and model stages consume power intermittently, while other modules draw continuous power. Practical results are based on measured energy usage from deployed components, whereas theoretical results represent lower bounds after applying the optimizations described in \Cref{sec:energy_consumption}.}
    }
    \label{tab:power_con_per_comp}
\end{table}

\subsubsection{Theoretical energy consumption model}
\label{sec:theoretical_power_consumption}
In the previous calculations, the degrees of freedom are limited by the development boards used. For practical deployments, energy consumption can be significantly reduced. 

(i) A first optimization is to deploy the transmitter and receiver on 1 platform. By doing this, only 1 \added[id=rev1]{$P_{plat}$} term remains in the energy calculation. To this end, different antenna designs should minimize self-interference between the transmitting and receiving antenna \cite{gordebeke_2024}.


(ii) Secondly, note that the transmitter and receiver make up 84~\% of the total power consumption. Part of this can be attributed to the fact that a \ac{dk} is used that is not fully optimized for energy consumption. Using the information stated in the datasheet of the DW3000 transceivers \cite{qorvo_dw3000_2020}, a theoretical minimal current profile can be calculated for both the transmitter and receiver. The current consumption of the transceiver will vary in time depending on the state in which the transceiver is. \Cref{form:theoretical_min_energy_TX} shows the calculation of the theoretical minimal energy consumption when the transceiver is in transmitter mode. \Cref{form:theoretical_min_energy_RX} calculates the used energy for the receiver. In these formulas, $E_{PLL}$ represents the energy consumed by the \ac{pll} phase, $E_{TX\_SHR}$ and $E_{RX\_SHR}$ denote the energy used while transmitting and receiving the \ac{shr}, and $E_{TX\_PHR\_PSDU}$ and $E_{RX\_PHR\_PSDU}$ correspond to the energy consumed during the transmission and reception of the \ac{phr} and \ac{psdu}. Additionally, $E_{PR\_{HUNT}}$ is the energy used during the preamble hunt phase, which involves the receiver searching for and detecting the preamble signal to synchronize with the incoming data transmission. $E_{RC}$ is the energy consumed in the receiver chain during the IDLE\_RC phase, and $E_{SLEEP}$ represents the energy used when the transceiver is in a low-power sleep state.

\begin{equation}
\label{form:theoretical_min_energy_TX}
\begin{split}
    E_{TX\_min} = &E_{PLL} + E_{TX\_SHR}\\
                  &+ E_{TX\_PHR\_PSDU} + E_{SLEEP}
\end{split}
\end{equation}

\begin{equation}
\label{form:theoretical_min_energy_RX}
\begin{split}
E_{RX\_min} = &E_{PLL} + E_{PR\_HUNT}\\
              &+ E_{RX\_SHR} + E_{RX\_PHR\_PSDU}\\
              &+ E_{PLL} + E_{RC} + E_{SLEEP}
\end{split}
\end{equation}

(iii) Thirdly, another major improvement can be made for the receivers' $E_{PR\_HUNT}$ term. The \ac{dk} receiver automatically reverts to the preamble hunt phase after receiving a packet. In the context of this solution, the timings between the transmitter and receiver are precisely defined. Also, as stated in optimization 1, both transmitter and receiver are deployed on a single platform, making it possible to share the same clock.
This enables us to minimize this preamble hunt phase, leading to a negligible phase hunt duration.

(iv) A final optimization concerns IDLE current usage of the transceivers. The development kit hardware lacks deep sleep capability and instead defaults to the \ac{pll} state, which consumes significantly more power. By implementing this functionality in the deployment version of the system, power consumption will be significantly reduced.



\begin{table}[h!]
\centering
\begin{tabular}{lllll}\hline
State & State name  & T (ms)        & I (mA)       & E (mJ)  \\\hline
\multicolumn{5}{c}{\textbf{Transmitter}}                         \\\hline
1        & PLL lock    & 0.020         & 18           & 0.00108  \\
2        & TX SHR      & 0.270         & 48           & 0.03888  \\
3        & TX PHR/PSDU & 0.016         & 40           & 0.0019   \\
4 (4Hz)  & SLEEP       & 249.694       & 0.00026      & 0.00019  \\
4 (20Hz) & SLEEP       &  49.694       & 0.00026      & 0.000039  \\\hline
\multicolumn{5}{c}{\textbf{Receiver}}                            \\\hline
1        & PLL lock    & 0.020         & 18           & 0.00108  \\
2        & PR\_HUNT    & 0.000         & 70           & 0           \\
3        & RX SHR      & 0.24          & 78           & 0.05616  \\
4        & RX PHR/PSDU & 0.016         & 70           & 0.00336  \\
5        & PLL lock    & 0.02          & 18           & 0.00108  \\
6        & IDLE RC     & 0.036         & 8            & 0.000864 \\
7 (4 Hz) & SLEEP       & 249.668       & 0.00026      & 0.00019  \\
7 (20 Hz)& SLEEP       &  49.668       & 0.00026      & 0.000039  \\
\end{tabular}
\caption{Theoretical duration and current usage of transmitter and receiver in the different states of the energy model.}
\label{tab:theoretical_current_and_timings}
\end{table}

Theoretical durations and current usages are presented in \Cref{tab:theoretical_current_and_timings}. Note that the transceiver operates at 3V, so all energy calculations were performed with this voltage in mind. The transmitter starts in the \ac{pll} state, using 18~mA for 0.020~ms. It then transmits the \ac{shr}, which is 256 symbols long, taking 0.270~ms and consuming 48~mA. Next, the \ac{phr} and \ac{psdu} are sent. For \ac{cir} collection, empty packets can be sent, but the \ac{psdu} will still contain the \ac{mac} header and \ac{rs} bits, taking 0.016~ms and consuming 40~mA. The transmitter then transitions into a deep sleep state for the rest of the 250 or 50-millisecond period, based on the configured sampling rate. In this case, 4 Hz and 20 Hz are shown. This state is very efficient, with a consumption of only 0.00026 mA. 
The receiver also starts in the \ac{pll} state and then skips the HUNT state. After this, the receiving device sequentially acquires the \ac{shr}, \ac{phr}, and \ac{psdu}, ending with another \ac{pll}, IDLE RC, and deep sleep state. 
When considering the system with a sampling rate of 20 Hz, 1 propagation sums to $0.0419~mJ$ for the transmitter and $0.0626~mJ$ for the receiver. \added[id=rev1]{Multiplying this by 20 gives the power consumption per second as stated in \Cref{tab:power_con_per_comp}.}
These new values give a new power consumption per 30 seconds: $E_{30\_sec\_min}=0.345~J$. Consequently, on average 0.0115~W is used, meaning the optimized system consumes one AAA's worth of energy ($5400~J$) every 130.5~h. A more realistic deployment option is a common battery pack of 20~000~mAh. This configuration could enable the system to work for 268.2~days without recharging. \added[id=rev1]{In less complex situations, where a sampling rate of 4 Hz is sufficient, $P_{TX\_min}$ and $P_{RX\_min}$ decrease even further. Consequently, $E_{30\_sec\_min}$ becomes $0.295~J$, enabling continuous monitoring for approximately 313.8 days.} This allows for \ac{br} monitoring in locations without power outlets, such as movable beds, furniture, and desks.

\section{Conclusion}
\label{sec:conclusion}

\added[id=rev1]{
This research developed and validated a non-intrusive system for accurate \ac{br} monitoring with potential applications in \added[id=rev2]{long-term remote patient monitoring, and presence detection in smart homes.} A \ac{cnn} model, based on an architecture specifically adapted for \ac{ir-uwb} \ac{cir} data and trained on a new open-source dataset, outperformed existing methods. Robustness was confirmed across distances up to 4~m (average L1 error 2.41~\ac{bpm}, reduced to 1.29~\ac{bpm} with calibration), body angles (0.86~\ac{bpm} at 45°), unseen environments, and a variety of postures.}

\added[id=rev1]{
Complexity analysis verified the preprocessing pipeline is lightweight, and the \ac{cnn} was effectively quantized with minimal accuracy loss, enabling deployment on a broader range of platforms. A novel analytical energy model was developed, showing that at 20~Hz the system consumes one AAA's worth of energy every 11.8~h, extending to every 130.5~h with theoretical optimizations. With a common 20~000~mAh battery pack, the optimized system operates for 268.2~days at 20~Hz, extending to 313.8~days at 4~Hz. Results also show that sampling at 4~Hz is sufficient in less complex scenarios, whereas a higher rate of 20~Hz is recommended for analyzing more complex cases.}

\added[id=rev1]{
Future research can address multi-person detection, \added[id=rev2]{non-static subjects, and the implementation of motion gating techniques to identify and filter out unreliable data during high-energy body movements. Exploring foundation models pre-trained on large-scale radar datasets could also enhance the system's ability to distinguish subtle respiratory patterns from complex body movements in non-static scenarios. Additionally, expanding the experimental scope to include higher respiratory rates and clinical populations would be essential for expanding the use case to acute respiratory distress scenarios. Finally, explainable \ac{ai} represents a highly promising direction, particularly in the context of healthcare applications.}
}


\bibliographystyle{IEEEtran}
\bibliography{reference.bib}

@online{eurostat2023,
    title = {Respiratory diseases statistics},
    author = {Eurostat},
    year = 2024,
    url = {https://ec.europa.eu/eurostat/statistics-explained/index.php?title=Respiratory_diseases_statistics},
    urldate = {01/04/2025}
}

@article{bousquet_impact_2013,
	title = {Impact of early diagnosis and control of chronic respiratory diseases on active and healthy ageing: {A} debate at the {European} {Union} {Parliament}},
	volume = {68},
	copyright = {http://onlinelibrary.wiley.com/termsAndConditions\#vor},
	issn = {0105-4538, 1398-9995},
	shorttitle = {Impact of early diagnosis and control of chronic respiratory diseases on active and healthy ageing},
	doi = {10.1111/all.12115},
	language = {en},
	number = {5},
	journal = {Allergy},
	author = {Bousquet, J. and Tanasescu, C.C. and Camuzat, T. and Anto, J.M. and Blasi, F. and Neou, A. and Palkonen, S. and Papadopoulos, N.G. and Antunes, J.P. and Samolinski, B. and Yiallouros, P. and Zuberbier, T.},
	month = may,
	year = {2013},
	pages = {555--561},
}

@article{kostikas_clinical_2020,
	title = {Clinical {Impact} and {Healthcare} {Resource} {Utilization} {Associated} with {Early} versus {Late} {COPD} {Diagnosis} in {Patients} from {UK} {CPRD} {Database}},
	volume = {15},
	issn = {1178-2005},
	doi = {10.2147/COPD.S255414},
	language = {eng},
	journal = {International Journal of Chronic Obstructive Pulmonary Disease},
	author = {Kostikas, Konstantinos and Price, David and Gutzwiller, Florian S. and Jones, Bethan and Loefroth, Emil and Clemens, Andreas and Fogel, Robert and Jones, Rupert and Cao, Hui},
	year = {2020},
	pmid = {32764917},
	pmcid = {PMC7371991},
	pages = {1729--1738},
}

@article{kayser_respiratory_2023,
	title = {Respiratory rate monitoring and early detection of deterioration practices},
	volume = {32},
	issn = {0966-0461, 2052-2819},
	doi = {10.12968/bjon.2023.32.13.620},
	language = {en},
	number = {13},
	journal = {British Journal of Nursing},
	author = {Kayser, Susan A and Williamson, Rachel and Siefert, Gabriela and Roberts, Dan and Murray, Angela},
	month = jul,
	year = {2023},
	pages = {620--627},
}

@INPROCEEDINGS{venkatesh_2005,
  author={Venkatesh, S. and Anderson, C.R. and Rivera, N.V. and Buehrer, R.M.},
  booktitle={MILCOM 2005 - 2005 IEEE Military Communications Conference}, 
  title={Implementation and analysis of respiration-rate estimation using impulse-based UWB}, 
  year={2005},
  volume={},
  number={},
  pages={3314-3320 Vol. 5},
  doi={10.1109/MILCOM.2005.1606167}}

@article{cheraghinia_comprehensive_2025,
	title = {A {Comprehensive} {Overview} on {UWB} {Radar}: {Applications}, {Standards}, {Signal} {Processing} {Techniques}, {Datasets}, {Radio} {Chips}, {Trends} and {Future} {Research} {Directions}},
	volume = {27},
	copyright = {https://ieeexplore.ieee.org/Xplorehelp/downloads/license-information/IEEE.html},
	issn = {1553-877X, 2373-745X},
	shorttitle = {A {Comprehensive} {Overview} on {UWB} {Radar}},
	url = {https://ieeexplore.ieee.org/document/10738385/},
	doi = {10.1109/COMST.2024.3488173},
	number = {4},
	urldate = {2026-05-15},
	journal = {IEEE Communications Surveys \& Tutorials},
	author = {Cheraghinia, Mohammad and Shahid, Adnan and Luchie, Stijn and Gordebeke, Gert-Jan and Caytan, Olivier and Fontaine, Jaron and Herbruggen, Ben Van and Lemey, Sam and Poorter, Eli De},
	month = aug,
	year = {2025},
	pages = {2283--2324},
}

@misc{plux_respiration_2020,
	title = {Respiration ({PZT}) Sensor Datasheet},
	url = {https://support.pluxbiosignals.com/wp-content/uploads/2021/11/Respiration_PZT_Datasheet.pdf},
	author = {{Plux}},
	urldate = {2025-04-01},
	date = {2020},
    year = {2020},
}

@online{idlab_idlab_2024,
	title = {{iDlab} Homelab},
	url = {https://homelab.ilabt.imec.be/},
	author = {{iDlab}},
	urldate = {2024-05-04},
	date = {2024},
    year = {2024},
}

@misc{qorvo_dw3000_2020,
	title = {{DW3000} {Datasheet}},
	url = {https://www.qorvo.com/products/d/da008142},
	urldate = {2025-02-14},
	author = {{Qorvo}},
	year = {2020},
}

@article{gordebeke_2024,
	title = {Time-{Domain}-{Optimized} {Antenna} {Array} for {High}-{Precision} {IR}-{UWB} {Localization} in {Harsh} {Urban} {Shipping} {Environments}},
	volume = {24},
	copyright = {https://ieeexplore.ieee.org/Xplorehelp/downloads/license-information/IEEE.html},
	issn = {1530-437X, 1558-1748, 2379-9153},
	doi = {10.1109/JSEN.2023.3310992},
	number = {5},
	journal = {IEEE Sensors Journal},
	author = {Gordebeke, Gert-Jan and Lemey, Sam and Caytan, Olivier and Boes, Michiel and Jocqué, Jelle and Van De Velde, Samuel and Marshall, Chris and De Poorter, Eli and Rogier, Hendrik},
	month = mar,
	year = {2024},
	pages = {5561--5577},
}

@article{RAJABY2022103403,
title = {A structured review of sparse fast Fourier transform algorithms},
journal = {Digital Signal Processing},
volume = {123},
pages = {103403},
year = {2022},
issn = {1051-2004},
doi = {https://doi.org/10.1016/j.dsp.2022.103403},
url = {https://www.sciencedirect.com/science/article/pii/S1051200422000203},
author = {Elias Rajaby and Sayed Masoud Sayedi}
}

@ARTICLE{Li_2024,
  author={Li, Qimeng and Liu, Jikui and Gravina, Raffaele and Zang, Weilin and Li, Ye and Fortino, Giancarlo},
  journal={IEEE Internet of Things Journal}, 
  title={A UWB-Radar-Based Adaptive Method for In-Home Monitoring of Elderly}, 
  year={2024},
  volume={11},
  number={4},
  pages={6241-6252},
  doi={10.1109/JIOT.2023.3310204}
}

@ARTICLE{Wang_2024,
  author={Wang, Pei and Ma, Xujun and Zheng, Rong and Chen, Luan and Zhang, Xiaolin and Zeghlache, Djamal and Zhang, Daqing},
  journal={IEEE Transactions on Mobile Computing}, 
  title={SlpRoF: Improving the Temporal Coverage and Robustness of RF-Based Vital Sign Monitoring During Sleep}, 
  year={2024},
  volume={23},
  number={7},
  pages={7848-7864},
  doi={10.1109/TMC.2023.3340925}
}

@INPROCEEDINGS{Kim_2024,
  author={Kim, Juhee and Kim, Seungku},
  booktitle={2024 IEEE International Conference on Consumer Electronics (ICCE)}, 
  title={Proposed Signal Processing Method for Continuous Respiration Monitoring using UWB Radar}, 
  year={2024},
  volume={},
  number={},
  pages={1-4},
  doi={10.1109/ICCE59016.2024.10444481}
}

@INPROCEEDINGS{Mu_2024,
  author={Mu, Wenyao and Zhang, Jinhui and Jiang, Xikang and Wang, Kun and Li, Lei and Zhang, Lin},
  booktitle={2024 IEEE International Workshop on Radio Frequency and Antenna Technologies (iWRF\&AT)}, 
  title={AP-HR: Amplitude and Phase Joint Optimization-Based Heartbeat and Respiration Separation Algorithm using IR-UWB Radar}, 
  year={2024},
  volume={},
  number={},
  pages={265-270},
  doi={10.1109/iWRFAT61200.2024.10594325}
}

@ARTICLE{Lopes_2022,
  author={Lopes, Alexandra and Osório, Daniel F. Noronha and Silva, Hugo and Gamboa, Hugo},
  journal={IEEE Sensors Journal}, 
  title={Equivalent Pipeline Processing for IR-UWB and FMCW Radar Comparison in Vital Signs Monitoring Applications}, 
  year={2022},
  volume={22},
  number={12},
  pages={12028-12035},
  doi={10.1109/JSEN.2022.3173218}
}

@INPROCEEDINGS{Zhou_2023,
  author={Zhou, Jialin},
  booktitle={2023 8th International Conference on Intelligent Computing and Signal Processing (ICSP)}, 
  title={MVSM: Motional Vital Signs Monitoring with IR-UWB Radar and Attention-Based CNN-LSTM Network}, 
  year={2023},
  volume={},
  number={},
  pages={679-683},
  doi={10.1109/ICSP58490.2023.10248662}
}

@Article{Dang_2022,
AUTHOR = {Dang, Xiaochao and Zhang, Jinlong and Hao, Zhanjun},
TITLE = {A Non-Contact Detection Method for Multi-Person Vital Signs Based on IR-UWB Radar},
JOURNAL = {Sensors},
VOLUME = {22},
YEAR = {2022},
NUMBER = {16},
ARTICLE-NUMBER = {6116},
PubMedID = {36015877},
ISSN = {1424-8220},
DOI = {10.3390/s22166116}
}

@Article{Leem_2017,
AUTHOR = {Leem, Seong Kyu and Khan, Faheem and Cho, Sung Ho},
TITLE = {Vital Sign Monitoring and Mobile Phone Usage Detection Using IR-UWB Radar for Intended Use in Car Crash Prevention},
JOURNAL = {Sensors},
VOLUME = {17},
YEAR = {2017},
NUMBER = {6},
ARTICLE-NUMBER = {1240},
PubMedID = {28556818},
ISSN = {1424-8220},
DOI = {10.3390/s17061240}
}

@Article{Khan_2017,
AUTHOR = {Khan, Faheem and Cho, Sung Ho},
TITLE = {A Detailed Algorithm for Vital Sign Monitoring of a Stationary/Non-Stationary Human through IR-UWB Radar},
JOURNAL = {Sensors},
VOLUME = {17},
YEAR = {2017},
NUMBER = {2},
ARTICLE-NUMBER = {290},
PubMedID = {28165416},
ISSN = {1424-8220},
DOI = {10.3390/s17020290}
}

@INPROCEEDINGS{Husaini_2022,
  author={Husaini, Muhammad and Kamarudin, Latifah Munirah and Zakaria, Ammar and Kamarudin, Intan Kartika and Ibrahim, Muhammad Amin},
  booktitle={2022 10th International Conference on Information and Communication Technology (ICoICT)}, 
  title={Non-Contact Vital Sign Monitoring During Sleep Through UWB Radar}, 
  year={2022},
  volume={},
  number={},
  pages={157-162},
  doi={10.1109/ICoICT55009.2022.9914876}
}

@ARTICLE{Zhang_2020,
  author={Zhang, Yi and Li, Xiuping and Qi, Rui and Qi, Zihang and Zhu, Hua},
  journal={IEEE Access}, 
  title={Harmonic Multiple Loop Detection (HMLD) Algorithm for Not-Contact Vital Sign Monitoring Based on Ultra-Wideband (UWB) Radar}, 
  year={2020},
  volume={8},
  number={},
  pages={38786-38793},
  doi={10.1109/ACCESS.2020.2976104}
}

@ARTICLE{Duan_2019,
  author={Duan, Zhenzhen and Liang, Jing},
  journal={IEEE Access}, 
  title={Non-Contact Detection of Vital Signs Using a UWB Radar Sensor}, 
  year={2019},
  volume={7},
  number={},
  pages={36888-36895},
  doi={10.1109/ACCESS.2018.2886825}
}

@Article{Wang_2020,
AUTHOR = {Wang, Dingyang and Yoo, Sungwon and Cho, Sung Ho},
TITLE = {Experimental Comparison of IR-UWB Radar and FMCW Radar for Vital Signs},
JOURNAL = {Sensors},
VOLUME = {20},
YEAR = {2020},
NUMBER = {22},
ARTICLE-NUMBER = {6695},
PubMedID = {33238557},
ISSN = {1424-8220},
DOI = {10.3390/s20226695}
}

@ARTICLE{Numan_2023,
  author={Numan, Paulson Eberechukwu and Park, Hyunwoo and Lee, Jaebok and Kim, Sunwoo},
  journal={IEEE Sensors Journal}, 
  title={Machine Learning-Based Joint Vital Signs and Occupancy Detection With IR-UWB Sensor}, 
  year={2023},
  volume={23},
  number={7},
  pages={7475-7482},
  doi={10.1109/JSEN.2023.3247728}
}

@INPROCEEDINGS{Farsaei_2023,
  author={Farsaei, Amirashkan and Meyer, Bernard and Sheikh, Alireza and El Soussi, Mohieddine and Zhang, Peng and Ramachandra, Gururaja Kasanadi and Govers, Jochem and Hijdra, Martijn},
  booktitle={2023 IEEE 34th Annual International Symposium on Personal, Indoor and Mobile Radio Communications (PIMRC)}, 
  title={An IEEE 802.15.4z-Compliant IR-UWB Radar System for In-Cabin Monitoring}, 
  year={2023},
  volume={},
  number={},
  pages={1-5},
  doi={10.1109/PIMRC56721.2023.10293800}
}

@article{choshenhillel_acute_2021,
	title = {Acute and chronic sleep deprivation in residents: {Cognition} and stress biomarkers},
	volume = {55},
	issn = {0308-0110, 1365-2923},
	shorttitle = {Acute and chronic sleep deprivation in residents},
	url = {https://asmepublications.onlinelibrary.wiley.com/doi/10.1111/medu.14296},
	doi = {10.1111/medu.14296},
	language = {en},
	number = {2},
	urldate = {2026-02-10},
	journal = {Medical Education},
	author = {Choshen‐Hillel, Shoham and Ishqer, Ahmad and Mahameed, Fadi and Reiter, Joel and Gozal, David and Gileles‐Hillel, Alex and Berger, Itai},
	month = feb,
	year = {2021},
	pages = {174--184},
}

@inproceedings{van2024wip,
  title={WiP paper: UWB-based integrated sensing and communication (ISAC) for robotic applications},
  author={Van Herbruggen, Ben and Luchie, Stijn and Jocqu{\'e}, Jelle and Wilssens, Ruben and Lemey, Sam and De Poorter, Eli},
  booktitle={EWSN2024, the International Conference on Embedded Wireless Systems and Networks},
  year={2024}
}

@article{coppens_uwb_2026,
	title = {{UWB} {TDoA} {Error} {Correction} {Using} {Transformers}: {Patching} and {Positional} {Encoding} {Strategies}},
	volume = {25},
	copyright = {https://ieeexplore.ieee.org/Xplorehelp/downloads/license-information/IEEE.html},
	issn = {1536-1276, 1558-2248},
	shorttitle = {{UWB} {TDoA} {Error} {Correction} {Using} {Transformers}},
	url = {https://ieeexplore.ieee.org/document/11237034/},
	doi = {10.1109/TWC.2025.3628168},
	urldate = {2026-05-15},
	journal = {IEEE Transactions on Wireless Communications},
	author = {Coppens, Dieter and Shahid, Adnan and De Poorter, Eli},
	year = {2026},
	pages = {7000--7013},
}

@incollection{porter_abnormal_2025,
	address = {Treasure Island (FL)},
	title = {Abnormal {Respirations}},
	copyright = {Copyright © 2025, StatPearls Publishing LLC.},
	url = {http://www.ncbi.nlm.nih.gov/books/NBK470309/},
	language = {eng},
	urldate = {2026-02-12},
	booktitle = {{StatPearls}},
	publisher = {StatPearls Publishing},
	author = {Porter, Rachel and Graham, Derrel D.},
	year = {2025},
	pmid = {29262235},
}





\end{document}